\documentclass[journal]{IEEEtran}
\IEEEoverridecommandlockouts
\usepackage[utf8]{inputenc}
\usepackage{bm}
\usepackage{amsfonts}
\usepackage{amssymb}
\usepackage{amsmath}
\usepackage{amsthm}
\usepackage{algorithm}
\usepackage{algpseudocode}
\usepackage{graphicx}
\usepackage{siunitx}
\usepackage{cite}
\usepackage{multicol}

\usepackage{lipsum}
\usepackage{color,soul}
\usepackage{comment}
\usepackage[dvipsnames]{xcolor}
\usepackage{pgfplots,tikz}
\usepackage[hidelinks]{hyperref}

\usetikzlibrary{spy}
\usetikzlibrary{external}
\pgfplotsset{compat=1.18}
\usetikzlibrary{arrows, shapes.geometric, shapes.misc, decorations, automata, backgrounds, petri, positioning, patterns}

\ifCLASSOPTIONpeerreview
\else
\fi

\title{Optimization of CF-mMIMO Systems for the Coexistence between eMBB+ and mMTC+: \\ From Analytical to GNN-Aided Designs}
\author{
Sergi~Liesegang,~\IEEEmembership{Member,~IEEE}, Lou~Salaün,~\IEEEmembership{Member,~IEEE}, \\ Chung~Shue~Chen,~\IEEEmembership{Senior Member,~IEEE}, and Stefano~Buzzi,~\IEEEmembership{Fellow,~IEEE}
    \thanks{A preliminary and compressed version of this paper has been presented in the conference paper \cite{Lie25a}.}
    \thanks{The European Union (EU) supported the work of S. Liesegang under the MSCA Postdoctoral Fellowship DIRACFEC (grant agreement No. 101108043). The EU also supported the work of S. Buzzi under the Italian National Recovery and Resilience Plan (NRRP) of NextGenerationEU, partnership on “Telecommunications of the Future” (PE00000001 - program “RESTART”, Structural Project 6GWINET, Cascade Call SPARKS). Views and opinions expressed are those of the authors and do not necessarily reflect those of the EU. The EU cannot be held responsible for them.}
    \thanks{S. Liesegang and S. Buzzi are with the Department	of Electrical and Information Engineering (DIEI), University of Cassino and Southern Latium (UNICAS), 03043 Cassino, Italy. S. Buzzi is also affiliated with the \textit{Consorzio Nazionale Interuniversitario per le Telecomunicazioni} (CNIT), 43124 Parma, Italy.}
    \thanks{L. Salaün and C.S. Chen are with Nokia Bell Labs, 91300 Massy, France.}
}
\begin{document}

\maketitle

\begin{abstract}
This paper investigates uplink multiple access for the coexistence of enhanced mobile broadband+ (eMBB+) and massive machine-type communications+ (mMTC+) in terminal-centric cell-free massive MIMO (CF-mMIMO) systems. We propose a non-orthogonal scheme in which low-rate mMTC+ transmissions are spread across the time-frequency grid shared with eMBB+ users, enabling efficient resource reuse. In the presence of imperfect channel state information, we derive closed-form expressions for the achievable rates of both services based solely on statistical channel knowledge. For mMTC+ devices, the analysis also incorporates finite blocklength (FBL) modeling to capture short-packet transmissions. To support heterogeneous service requirements, we formulate a power-control problem that maximizes the minimum energy efficiency of mMTC+ devices subject to quality-of-service constraints on eMBB+ users. The resulting nonconvex problem is solved via sequential fractional programming, accounting for both the Shannon and FBL regimes. To enable real-time operation, we further propose a graph neural network (GNN) with multi-head attention to approximate the model-based solution. Constraint satisfaction during training is enforced via an augmented Lagrangian loss. Numerical results demonstrate effective multiplexing of the two data services and show that the proposed GNN algorithm achieves near-optimal performance with a significantly lower computational complexity.
\end{abstract}


\section{Introduction} \label{sec:1}
In this decade, both academia and industry are actively developing sixth-generation cellular networks (6G) \cite{Hua22}. The exponential rise in the number of connected devices, together with the ever-growing demand for heterogeneous data services, has revealed the need for new communication solutions \cite{Eri23}. 6G will build upon the use cases of previous generations: enhanced mobile broadband+ (eMBB+), targeting large data rates; ultra-reliable low-latency communications+ (URLLC+), which require short delays and high reliability; and massive machine-type communications+ (mMTC+), characterized by massive connectivity and low consumption. The diversity of these requirements makes it challenging to support all pillars simultaneously.

Consequently, coexistence will play a crucial role in the design and implementation of future mobile networks. Emerging technologies such as millimeter-wave bands, centralized and distributed large-scale multiple-input multiple-output (MIMO), reconfigurable intelligent surfaces, and edge intelligence are promising solutions to the above challenges \cite{Ngu22}.

In earlier releases, deploying a large number of antennas at (macro) base stations, namely massive MIMO (mMIMO) \cite{Ozd19}, has enabled operators to increase achievable data rates significantly. However, the associated hardware complexity and power consumption scale with the number of antennas, which soon becomes prohibitive. Furthermore, these centralized architectures often fail to provide satisfactory quality of service (QoS) to users located far from the base station, mainly due to poor channel conditions and inter-cell interference.

To address these issues, a compelling alternative is to deploy the antennas across the coverage area as access points (APs) and adopt a \textit{user-centric} network design \cite{Dem21}. This leads to a distributed architecture without cell borders known as \textit{cell-free} massive MIMO (CF-mMIMO), which has been shown to improve QoS and reduce power consumption compared to its collocated counterparts \cite{BuzziWhy}.

Multi-antenna technologies also facilitate the deployment of heterogeneous networks, as the resulting spatial diversity can be exploited to mitigate interference among different data services. However, the coexistence of human-centric and machine-type communications is often addressed using orthogonal multiple access (MA) schemes, which may be outperformed by non-orthogonal approaches \cite{Jor24, Moh24}.

Building on the previous discussion and drawing on the findings reported in \cite{Lie25a}, we investigate MA strategies to mitigate inter-service interference. In particular, this work considers an uplink (UL) cell-free (CF) scenario in which eMBB+ users and mMTC+ devices share resources. Since mMTC+ services typically involve low data rates, we propose to embed their transmissions as spread-spectrum signals \cite{Sim94} within the time-frequency resources allocated to eMBB+ users. Owing to the low transmit power of the (long) spread mMTC+ packets, this MA strategy can substantially reduce interference among the two heterogeneous services \cite{Mol22}.

In this paper, we also consider imperfect channel state information (CSI). Specifically, we derive closed-form lower bounds on achievable data rates that rely solely on statistical channel knowledge. It is worth noting that the decorrelation of the mMTC+ signature sequences leads to (cumbersome) $8$-th order statistical moments, for which explicit analytical expressions are obtained. For performance evaluation, we subsequently address the design of power control strategies under a fairness-oriented policy, aiming to maximize the energy efficiency (EE) of mMTC+ transmissions \cite{Buz16} while enforcing, among others, QoS constraints on data rates for eMBB+ users. Besides, given the short-packet transmissions of mMTC+ applications, e.g., narrowband IoT (NB-IoT), we also incorporate the finite blocklength (FBL) regime into our analysis \cite{Pol10}. To the best of our knowledge, results of this kind have not been previously reported in the literature.

Owing to the signal model's characteristics, the resulting power-control optimization problems are inherently nonconvex (cf. \cite{Dem21}). Consequently, they can only be handled through iterative optimization techniques such as \textit{sequential} fractional programming (FP) \cite{Mat20}. However, these approaches entail very high computational complexity that grows rapidly with network size, raising concerns about their suitability for real-time implementation \cite{Hoy21}. This observation strongly motivates exploring alternative methods that maintain good performance while reducing computational burden.

To overcome this limitation, we will employ machine learning (ML) tools and recast the problem in terms of graph neural networks (GNNs) \cite{Guo22}, whose structure naturally aligns with CF-mMIMO deployments. Moreover, GNNs have been shown to efficiently exploit graph topology \cite{She23}, endowing them with inherent generalization properties. In practice, this allows them to operate in scenarios with varying numbers of users and/or APs without requiring retraining \cite{Mis24}. In this framework, analytical (or model-based) solutions will be used to generate extensive datasets for training GNNs.

Unfortunately, predictions from conventional GNN training are not guaranteed to satisfy all original constraints \cite{Eis19}. In some cases, simple post-processing is enough to ensure feasibility (e.g., linear scaling). However, this only applies to plain constraints (e.g., power budgets). When dealing with more complex constraints (e.g., QoS), such transformations are no longer trivial. This is why we introduce additional terms in the learning loss to enforce that the output satisfies the constraints. As discussed later in this paper, we advocate the augmented Lagrangian method, which penalizes the model whenever the constraints are violated \cite{Boy11}.

Finally, all the supervised ML approaches are benchmarked against their theoretical counterparts, assessing their performance with respect to (w.r.t.) EE, data rate, and computational complexity. This comparison enables us to quantify the advantages of ML-based solutions in scenarios where sustainability and fairness requirements must be satisfied simultaneously, hence addressing a design problem that remains largely unexplored yet is of growing importance for 6G systems.

\subsection{State of the Art}
In the early stages of 5G, coexistence was primarily addressed via \textit{network slicing}, in which services were allocated orthogonal resources. This perspective has gradually shifted toward non-orthogonal alternatives (e.g., \cite{Pop18}). For example, \cite{Liu24} analyzes different slicing schemes for heterogeneous services and shows that rate-splitting MA yields greater flexibility, outperforming orthogonal and non-orthogonal MAs by achieving higher rates for users with diverse service requirements. The co-scheduling of eMBB and URLLC traffic is also investigated in \cite{Bai21} using a puncturing-based resource sharing scheme. The authors formulate a maximin rate optimization problem for eMBB users subject to URLLC constraints and propose slot- and mini-slot-level scheduling algorithms that achieve improved fairness and performance over baseline methods. In \cite{Int23}, the coexistence of eMBB and URLLC in the downlink (DL) of multi-cell (MC) mMIMO systems is studied, and puncturing and superposition coding strategies are compared under imperfect CSI. Results indicate that superposition coding generally achieves higher eMBB spectral efficiency (SE) while meeting URLLC reliability targets, whereas puncturing may be preferable in high-interference scenarios. Finally, the coexistence of eMBB and URLLC in scalable CF-mMIMO systems with hardware impairments is examined in \cite{Fem26}, showing that practical transceiver non-idealities degrade channel estimation, thereby reducing eMBB performance and URLLC reliability.

ML for wireless communications has also attracted considerable attention in recent years (see \cite{Bjo20b}). Theoretical foundations for applying GNNs to wireless network optimization are presented in \cite{She23}, demonstrating that they can achieve near-optimal performance with significantly fewer training samples than conventional neural networks and exhibit strong scalability and generalization across diverse network settings. In \cite{Guo22}, the authors propose a heterogeneous GNN framework for power allocation in MC multi-user systems that exploits the policy's permutation equivariance. By incorporating appropriate parameter sharing, the method improves learning efficiency and achieves sum-rate performance comparable to conventional deep neural networks, while substantially reducing training and computational complexity. In \cite{Nad23}, a GNN-based approach to joint user selection and power control in interference networks is proposed, in which resource allocation is formulated as a primal-dual optimization problem with adaptive minimum-capacity constraints. The resulting permutation-equivariant GNN learns scalable policies that improve fairness-rate trade-offs compared to baseline algorithms. Finally, \cite{Mis24} investigates DL power allocation in CF-mMIMO systems and proposes a GNN that maximizes the minimum signal-to-interference-plus-noise ratio (SINR), achieving near-optimal SE with significantly lower computational complexity than the second-order cone programming solution.

Despite notable progress, most of these papers focus on the DL and on the coexistence of eMBB and URLLC. To the best of our knowledge, no prior studies have reported (real-time) power-control design for scalable CF-mMIMO networks with eMBB+ and mMTC+ in the UL under QoS constraints.

\subsection{Contributions} \label{sec:1.2}
The purpose of this paper, which can be cast as an extension of the preliminary results presented by the same authors in the conference paper \cite{Lie25a}, is to fill the previous gaps in the existing literature. More precisely, the contributions of this work are:
\begin{enumerate}
    \item We propose novel non-orthogonal MA schemes that enable the coexistence of eMBB+ and mMTC+ in scalable CF-mMIMO systems (in which terminals are only served by a few APs). Since mMTC+ devices operate at low data rates, their signals are spread across the time-frequency resource grid shared with eMBB+.
    \item We derive closed-form analytical expressions for the achievable rates of eMBB+ users and mMTC+ devices, based solely on statistical information. To represent the short-packet transmissions of mMTC+, the FBL analysis is also applied to these terminals.
    \item We design power-control policies that maximize the minimum EE among mMTC+ devices, subject to QoS constraints on eMBB+ users. This way, we account for the different requirements of the two 6G services (power consumption and data rate). 
    \item We solve the optimization via sequential FP, which guarantees convergence to stationary points (local optima of the problem). This is first done using classical Shannon's analysis and later under the FBL regime.
    \item We propose a GNN-based implementation with multi-head attention to approximate the theoretical approach, thereby reducing complexity and paving the way for real-time applications. Also, to ensure that all optimization constraints are satisfied during training, we modify the loss using the augmented Lagrangian method.
\end{enumerate}

The study is completed with extensive numerical experiments that evaluate the effectiveness of the proposed designs in terms of EE, data rate, and computational complexity. The simulations demonstrate the satisfactory performance of both algorithms. These outcomes also allow us to elucidate the crucial role of expert knowledge in the learning process and to reveal the GNN's generalization capabilities.

\subsection{Organization} \label{sec:1.3}
This paper is structured as follows. Section~\ref{sec:2} introduces the system model. Section~\ref{sec:3} derives the achievable data rates. Section~\ref{sec:4} formulates the optimization problem, and the analytical solution is given in Section~\ref{sec:5}. Section~\ref{sec:6} presents the GNN-aided implementation. Section~\ref{sec:7} is devoted to the numerical experiments. Section~\ref{sec:8} concludes the work.

\subsection{Notation} \label{sec:1.4}
In this work, scalars, vectors, and matrices are denoted by italic, boldface lower-case, and upper-case letters, respectively. $\mathbf{0}_m$ denotes the all-zeros vector of length $m$ and $\mathbf{I}_m$ denotes the identity matrix of size $m \times m$. $\mathbb{R}^{m \times n}$ and $\mathbb{C}^{m \times n}$ denote the $m$ by $n$ dimensional real and complex spaces, respectively. The transpose, Hermitian, inverse, trace, and expectation operators are denoted by $(\cdot)^{\mathrm{T}}$, $(\cdot)^{\mathrm{H}}$, $(\cdot)^{-1}$, $\mathrm{tr}(\cdot)$, and $\mathbb{E}[\cdot]$ respectively. The sign function is denoted $\mathrm{sgn}(\cdot)$. $\mathcal{CN}(\cdot,\cdot)$ denotes the complex proper Gaussian distribution.

\section{System Model} \label{sec:2}
\begin{figure}[t]
\centerline{\includegraphics[scale=0.175]{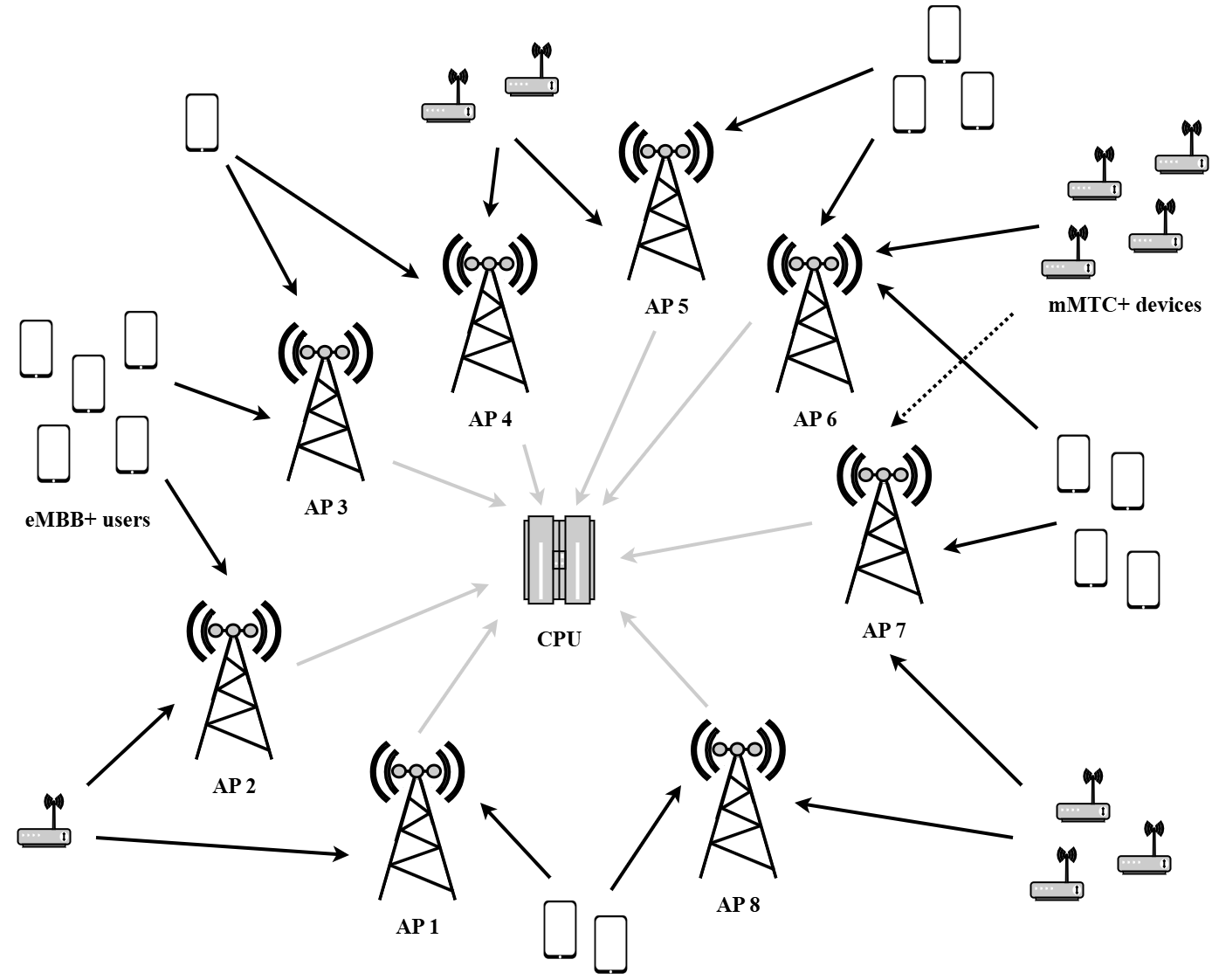}}
\vspace{-2mm}
\caption{Illustrative example of a \textit{terminal-centric} CF-mMIMO setup, where subsets of $M_s = 2$ APs (out of $M = 8$), equipped with $L = 3$ antennas, serve $K_u = 15$ eMBB+ users and $K_d = 10$ mMTC+ devices simultaneously.}
\label{fig:1}
\vspace{-2mm}
\end{figure}

In this work, we consider a CF-mMIMO deployment similar to the one presented in \cite{Lie25a}, where $M$ APs, each equipped with $L$ antennas, are connected to a central processing unit (CPU) through high-capacity fronthaul links and jointly serve $K_u$ single-antenna eMBB+ users and $K_d$ single-antenna mMTC+ devices. This system model is depicted in Fig.~\ref{fig:1}, where it can be observed that the different terminals\footnote{For brevity, throughout this paper, the term “terminals” is used interchangeably to refer to both eMBB+ users and mMTC+ devices.} communicate with only a subset of $M_s \leq M$ available APs \cite{Elw23}. In this sense, the conventional user-centric paradigm naturally extends to a \textit{terminal-centric} architecture.

Recall that both users and devices\footnote{For further conciseness, the terms “users” and “devices” will exclusively denote eMBB+ and mMTC+ terminals, respectively.} operate over the same set of resources. Specifically, following the 3GPP terminology \cite{3GPP36814}, we model the system using a grid composed of $N$ physical resource blocks (PRBs) in the time-frequency domain. To enable the coexistence of the two services, a spread-spectrum MA scheme is adopted for mMTC+ devices \cite{Sim94}. This choice facilitates message separation and, as discussed later, helps reduce the interference experienced by eMBB+ users. Moreover, unlike eMBB+ transmissions, which typically demand high data rates, mMTC+ devices are generally energy-limited and therefore require highly energy-efficient operation. That is, the resulting \textit{spreading gain} can be exploited to lower peak transmit power levels \cite{Mol22}.

This is presented in the sequel, where the two stages in the UL, namely channel estimation and data transmission, are described in detail. Before that, we devote a subsection to characterizing the propagation model.

\subsection{Propagation Channel}\label{sec:2.1}
Assuming channel stationarity over the resource grid, the link from user $u$ to AP $m$ at PRB $n$ is \cite{Int23}
\begin{equation}
    \mathbf{h}_{m,u}[n] \sim \mathcal{CN}(\mathbf{0},\mathbf{R}_{m,u}),
    \label{eq:1} 
\end{equation}
where $\mathbf{R}_{m,u} \in \mathbb{C}^{L \times L}$ refers to the spatial correlation matrix of the Rayleigh-distributed non-line-of-sight (NLoS) components. The corresponding large-scale fading (LSF) coefficient, encompassing path loss, is denoted by $\alpha_{m,u} = \mathrm{tr}(\mathbf{R}_{m,u})/L$. 

For the device-AP link, we also adopt a Rayleigh model:
\begin{equation}
    \mathbf{g}_{m,d}[n] \sim \mathcal{CN}(\mathbf{0},\mathbf{Q}_{m,d}),
    \label{eq:2} 
\end{equation}
with $\mathbf{Q}_{m,d} \in \mathbb{C}^{L \times L}$ the correlation in the NLoS propagation and $\beta_{m,d} = \mathrm{tr}(\mathbf{Q}_{m,d})/L$ the set of LSF parameters.

\subsection{Uplink Channel Estimation}\label{sec:2.2}
Assuming perfect CSI is often overly optimistic for practical systems. Instead, it is more realistic to acquire this information locally at the APs by means of UL orthogonal pilot transmissions\footnote{In practical deployments, the low activity of mMTC+ terminals allows for pilot-based channel estimation, as the number of active devices is comparable to that of eMBB+ users.}. This procedure enables the characterization of the sufficient statistics of the channel estimates \cite[(5)]{Ozd19}.

A feasible option could be the minimum mean-square error (MMSE) estimation \cite[Subsection~V-B]{Lie25a}:
\begin{equation}
    \hat{\mathbf{h}}_{m,u}[n] = \sqrt{\eta_u} \mathbf{A}_{m,u} \bar{\mathbf{h}}_{m,u}[n],
    \label{eq:3}
\end{equation}
where $\eta_u$ ($\zeta_d$) is the training power, $\mathbf{A}_{m,u} \triangleq \mathbf{R}_{m,u} \mathbf{C}_{m,u}^{-1}$ refers to the MMSE matrix, and
\begin{equation}    
    \mathbf{C}_{m,u} = \sum_{k = 1}^{K_u} \eta_k \mathbf{R}_{m,k} \left\vert \bm{\phi}_k^{\mathrm{H}} \bm{\phi}_u \right\vert^2 + \sum_{d = 1}^{K_d} \zeta_d \mathbf{Q}_{m,d} \left\vert \bm{\pi}_d^{\mathrm{H}} \bm{\phi}_u \right\vert^2 + \sigma_m^2 \mathbf{I}_L,        
    \label{eq:4}
\end{equation}
denotes the covariance matrix of the least-squares observations $\bar{\mathbf{h}}_{m,u} [n] \in \mathbb{C}^L$ providing sufficient statistics, i.e.,
\begin{equation}
    \begin{aligned}
        \bar{\mathbf{h}}_{m,u}[n] &= \sqrt{\eta_u} \mathbf{h}_{m,u}[n] + \sum\nolimits_{k \neq u} \sqrt{\eta_k} \mathbf{h}_{m,k}[n] \bm{\phi}_k^{\mathrm{H}} \bm{\phi}_u \\
        &\quad +  \sum\nolimits_{d = 1}^{K_d} \sqrt{\zeta_d} \mathbf{g}_{m,d}[n] \bm{\pi}_d^{\mathrm{H}} \bm{\phi}_u + \bm{\omega}_{m,u} [n],
    \end{aligned}
    \label{eq:5}
\end{equation}
with $\bm{\phi}_u \in \mathbb{C}^{\tau_p}$ ($\bm{\pi}_d \in \mathbb{C}^{\tau_p}$) the sequence with length $\tau_p$ pilots sent by user $u$ (device $d$) to estimate the channel in all PRBs; and $\bm{\omega}_{m,u} [n] \in \mathbb{C}^L$ the ambient noise with variance $\sigma_m^2$. Please see \cite[Subsection~II-B]{Int23} for more details. 

The expressions for the mMTC+ estimates can be obtained similarly, yet are not included to avoid redundancy.

\subsection{Uplink Data Transmission} \label{sec:2.3}
Accordingly, the signal received at the $m$-th AP reads as
\begin{equation}    
    \mathbf{r}_m[n] =  \mathbf{y}_m^{\text{eMBB+}}[n] + \mathbf{y}_m^{\text{mMTC+}}[n] + \mathbf{w}_m[n],        
    \label{eq:6}
\end{equation}
where $\mathbf{w}_m[n]$ represents the additive white Gaussian noise, i.e., $\mathbf{w}_m[n] \sim \mathcal{CN}(\mathbf{0}_L,\sigma_m^2 \mathbf{I}_L)$.

The first term in \eqref{eq:6} contains the information of the eMBB+ terminals and can be modeled as follows:
\begin{equation}    
    \mathbf{y}_m^{\text{eMBB+}} [n] = \sum\nolimits_{u = 1}^{K_u} \sqrt{p_u}\mathbf{h}_{m,u}[n]s_u[n],            
    \label{eq:7}
\end{equation}
where $p_u$ is the UL power budget, $\mathbf{h}_{m,u}[n] \in \mathbb{C}^L$ is the user-AP channel from \eqref{eq:1}, and $s_u[n]$ is the transmit signal of user $u$. We also consider that all $s_u[n]$ are normally distributed with zero mean and unit power, i.e., $s_u[n] \sim \mathcal{CN}(0,1)$.

The second term comprises the mMTC+ messages, i.e.,
\begin{equation}
    \mathbf{y}_m^{\text{mMTC+}} [n] = \sum\nolimits_{d = 1}^{K_d} \sqrt{q_d} \mathbf{g}_{m,d}[n] x_d[n],
    \label{eq:8}
\end{equation}
where $q_d$ is the power coefficient and $\mathbf{g}_{m,d}[n]$ is the device-AP channel. In that sense, $x_d[n]$ is the signal transmitted by device $d$ and can be expressed as
\begin{equation}
    x_d[n] = c_d[n] z_d,
    \label{eq:9}
\end{equation}
where $c_d[n] \in \mathbb{C}$ is the unit-energy spreading waveform generated from a \textit{pseudo-noise} (PN) sequence of length $W$ and cross-correlation values equal to $1/W$ \cite{Mol22}. In this work, we assume that the spreading spans both time and frequency simultaneously, so that $W = N$ (i.e., all available PRBs). Thus, the data symbol $z_d$ is shared across all time-frequency resources. As before, $z_d$ is modeled as a standard complex Gaussian random variable: $z_d \sim \mathcal{CN}(0,1)$.

\subsubsection{Spatial Detection}
By applying linear receive combiners $\mathbf{f}_{m,u}[n], \mathbf{t}_{m,d}[n] \in \mathbb{C}^L$, each AP detects the signals of its associated terminals and forwards the corresponding outputs to the CPU, where the signals $s_u[n]$ and $x_d[n]$ are recovered. As previously noted, in practical and scalable CF deployments, only a subset of APs serves each user and device \cite{Bjo20a}. For notational convenience, this association is represented by the binary coefficients $a_{m,u}$ and $b_{m,d}$, which are $1$ when the terminal and the AP are connected.

Under this model, the eMBB+ information is obtained directly via aggregation, i.e., $\hat{s}_u[n] = \sum_{m=1}^M a_{m,u} \mathbf{f}_{m,u}^{\mathrm{H}}[n] \mathbf{r}_m[n]$, one for each time-frequency slot. In contrast, due to the use of PN sequences for mMTC+ transmissions, additional processing is required to recover the underlying symbols $z_d$.

\subsubsection{Time-Frequency Despreading}
Once the received signal is equalized utilizing filters $\mathbf{t}_{m,d}[n]$, each AP will stack all its replicas $\tilde{x}_{m,d}[n] = \mathbf{t}_{m,d}^{\mathrm{H}}[n] \mathbf{r}_m[n]$ into a large column vector $\tilde{\mathbf{x}}_{m,d} = [\tilde{x}_{m,d}[1],\ldots,\tilde{x}_{m,d}[N]]^{\mathrm{T}}$ for later correlating it with the (modified) signature PN sequences, i.e.,
\begin{equation}
    \tilde{z}_{m,d} = \mathbf{c}_d^{\mathrm{H}} \mathrm{diag}\left(\tilde{\mathbf{g}}_{m,d}\right) \tilde{\mathbf{x}}_{m,d},
    \label{eq:10}
\end{equation}
where $\mathbf{c}_d = [c_d[1],\ldots,c_d[N]]^{\mathrm{T}}$ is the original spreading waveform and the vector $\tilde{\mathbf{g}}_{m,d} = [\tilde{g}_{m,d}[1],\ldots,\tilde{g}_{m,d}[N]]^{\mathrm{T}}$ concatenates the set of processed gains
\begin{equation}
    \tilde{g}_{m,d}[n] = \mathbf{g}_{m,d}^{\mathrm{H}}[n] \mathbf{t}_{m,d}[n].
    \label{eq:11}
\end{equation}

To some extent, \eqref{eq:10} may be interpreted as a \textit{time-frequency despreading} operation \cite{Sim94}. However, unlike in single-antenna systems, the weights $\tilde{g}_{m,d}[n]$ are required to account for the spatial diversity gains. Otherwise, the new \textit{degrees of freedom} may impair the spreading gain, as the (auto-/cross-) correlation properties of the PN sequences would no longer be preserved.

To further mitigate mMTC+ interference, \eqref{eq:10} can be designed according to a zero-forcing (ZF) criterion. In short, instead of directly applying the sequence $\bm{c}_d$ (which resembles a matched filter), the correlating sequence of device $d$ may be projected onto the null space spanned by the other signatures \cite{Sim94}. A detailed analysis of this approach is beyond the scope of this work and will be addressed in future research.

Note that, regardless of the despreading operation, the CPU will ultimately obtain the estimates $\hat{z}_d = \sum_{m=1}^M b_{m,d} \tilde{z}_{m,d}$, which will then be used to decode the mMTC+ data.

\section{Achievable Data Rate} \label{sec:3}
In the following section, we formulate the power control problem to maximize the minimum EE among the mMTC+ devices while enforcing QoS constraints for both terminal types. To this end, we first derive an achievable lower bound on the data rates for both terminals under imperfect CSI. The key distinction arises from the use of short packets, which are characteristic of mMTC+ transmissions and necessitate analysis within the FBL regime.

\subsection{eMBB+} \label{sec:3.1}
Under imperfect CSI, a tractable expression for the achievable data rate can be obtained by using the hardening or \textit{use-and-then-forget} (UatF) bound, which is commonly adopted in mMIMO analyses. In essence, this approach uses the channel estimates solely for beamforming and then discards them during signal detection \cite{Int23}. More precisely, considering that the CPU has access only to channel distribution information (CDI), but not to the instantaneous channel realizations, the received signal of user $u$ can be written as in \eqref{eq:12} at the top of the next page, where the estimation errors are modeled as an additional “effective” noise term $U_u[n]$. In line with the reasoning in \cite[Theorem 5.4]{Dem21}, we consider the worst-case setting and represent $U_u[n]$ as uncorrelated Gaussian noise, which yields a lower bound on the (ergodic) capacity. This way, the achievable data rate of user $u$ in bits/Hz (or SE) reads as
\begin{equation}
    R_u^{\text{eMBB+}}(\gamma_u) = \log_2\left(1 + \gamma_u\right).
    \label{eq:13}
\end{equation}
and the SINR is
\begin{equation}
    \gamma_u = \frac{\displaystyle \delta_u p_u}{\displaystyle \upsilon_u p_u + \sum\nolimits_{k \neq u} \kappa_{u,k} p_k + \sum\nolimits_d \varkappa_{u,d} q_d + \xi_u},
    \label{eq:14}
\end{equation}
where $\delta_u = \left\vert D_u \right\vert^2$ is the strength of the useful signal and $\upsilon_u = \mathbb{E}[\vert U_u[n] \vert^2]$ represents the channel uncertainty, whilst $\kappa_{u,k} = \mathbb{E}[\vert I_{u,k}^{\text{eMBB+}} [n] \vert^2]$, $\varkappa_{u,d} = \mathbb{E}[\vert I_{u,d}^{\text{mMTC+}} [n]\vert^2]$, and $\xi_u = \mathbb{E}[\vert W_u^{\text{eMBB+}}[n] \vert^2 ]$ refer to the powers of the (eMBB+/mMTC+) interference and noise, respectively. As shown, the key idea is that the mMTC+ contribution is nearly negligible, since the devices' transmit powers $q_d$ are significantly smaller than the coefficients $p_u$ associated with the eMBB+ users.

\begin{figure*}[t]
\begin{align}     
    \hat{s}_u [n] &= \sqrt{p_u} \underbrace{\mathbb{E}\left[\sum_m a_{m,u} \mathbf{f}_{m,u}^{\mathrm{H}}[n] \mathbf{h}_{m,u}[n] \right]}_{\triangleq D_u}s_u[n] + \sqrt{p_u} \underbrace{\left(\sum_m a_{m,u} \mathbf{f}_{m,u}^{\mathrm{H}}[n] \mathbf{h}_{m,u}[n] - D_u\right)}_{\triangleq U_u[n]} s_u[n] + \underbrace{\sum_m a_{m,u} \mathbf{f}_{m,u}^{\mathrm{H}}[n]\mathbf{w}_m[n]}_{\triangleq W_u^{\text{eMBB+}}[n]} \nonumber \\
    &\quad + \sum_{k \neq u} \sqrt{p_k} \underbrace{\sum\nolimits_m a_{m,u} \mathbf{f}_{m,u}^{\mathrm{H}}[n] \mathbf{h}_{m,k}[n]}_{\triangleq I_{u,k}^{\text{eMBB+}}[n]} s_k[n]  + \sum_d \sqrt{q_d}  \underbrace{\sum\nolimits_m a_{m,u} \mathbf{f}_{m,u}^{\mathrm{H}}[n] \mathbf{g}_{m,d}[n]}_{\triangleq I_{u,d}^{\text{mMTC+}}[n]} x_d[n],    
    \label{eq:12}
\end{align}
\vspace{-2mm}
\hrule
\end{figure*}

Deriving closed-form expressions for the moments above is generally challenging. However, under maximum-ratio combining (MRC)\footnote{Note that this work can also be extended to more sophisticated processing schemes such as ZF and MMSE. However, instead of analytical expressions (available under MRC), the statistical moments would be computed numerically. The interested reader is referred to \cite{Bjo20a} and the references therein.}, i.e., $\mathbf{f}_{m,u}[n] = \hat{\mathbf{h}}_{m,u}[n]$ (and $\mathbf{t}_{m,u}[n] = \hat{\mathbf{g}}_{m,u}[n]$), they can be obtained after suitable manipulations. In particular, the following result holds \cite[Corollary~2]{Bjo20a}:
\begin{equation}
    \begin{aligned}
        \delta_u &= \eta_u^2 \left\vert \sum\nolimits_m a_{m,u} \mathrm{tr}\left(\mathbf{A}_{m,u} \mathbf{R}_{m,u}\right) \right\vert^2, \\        
        \upsilon_u &= \eta_u \sum\nolimits_m a_{m,u} \mathrm{tr}\left(\mathbf{A}_{m,u} \mathbf{R}_{m,u} \mathbf{R}_{m,u}\right), \\
        \kappa_{u,k} &= \eta_u \sum\nolimits_m a_{m,u} \mathrm{tr}\left(\mathbf{A}_{m,u} \mathbf{R}_{m,u} \mathbf{R}_{m,k}\right) \\
        &\quad + \eta_u \eta_k \left\vert \bm{\phi}_k^{\mathrm{H}} \bm{\phi}_u \right\vert^2 \left\vert \sum\nolimits_m a_{m,u} \mathrm{tr}\left(\mathbf{A}_{m,u} \mathbf{R}_{m,k}\right) \right\vert^2, \\
    \end{aligned}
    \label{eq:15}
\end{equation}

\begin{equation*}
    \begin{aligned}
        \varkappa_{u,d} &= \eta_u \sum\nolimits_m a_{m,u} \mathrm{tr}\left(\mathbf{A}_{m,u} \mathbf{R}_{m,u} \mathbf{Q}_{m,d}\right) \\
        &\quad + \eta_u \zeta_d \left\vert \bm{\pi}_d^{\mathrm{H}} \bm{\phi}_u \right\vert^2 \left\vert \sum\nolimits_m a_{m,u} \mathrm{tr}\left(\mathbf{A}_{m,u} \mathbf{Q}_{m,d}\right) \right\vert^2, \\
        \xi_u &= \eta_u \sum\nolimits_m a_{m,u} \sigma_m^2 \mathrm{tr}\left(\mathbf{A}_{m,u} \mathbf{R}_{m,u}\right). \phantom{\eta_u \sum\nolimits_m a_{m,u} \sigma_m^2 \mathrm{tr}\left(\mathbf{A}_{m,u} \mathbf{R}_{m,u}\right).}
    \end{aligned}
\end{equation*}

\subsection{mMTC+} \label{sec:3.2}
To derive the UatF bound for device $d$, we first rewrite the received signal as in \eqref{eq:16}, reported at the top of page 7, where $\tilde{\mathbf{t}}_{m,d}[n] = \mathbf{g}_{m,d}^{\mathrm{H}}[n] \mathbf{t}_{m,d}[n] \mathbf{t}_{m,d}^{\mathrm{H}}[n]$. As before, the effect of imperfect CSI is incorporated through an effective noise term $V_u$. It is worth noting that, unlike \eqref{eq:12}, where each user observes $N$ distinct realizations across the resource grid, the expression in \eqref{eq:16} is common to each device (see also \eqref{eq:10}).

As a result, we can show that the SINR yields
\setcounter{equation}{16}
\begin{equation}
    \rho_d = \frac{\displaystyle \lambda_d q_d}{\displaystyle \nu_d q_d + \sum\nolimits_{k \neq d} \epsilon_{d,k} q_k + \sum\nolimits_u \varepsilon_{d,u} p_u + \chi_d},
    \label{eq:17}
\end{equation}
where $\lambda_d = \vert S_d \vert^2$ denotes the power of the desired signal, $\nu_d = \mathbb{E}[\vert V_d \vert^2]$ comprises the effect of imperfect CSI, and $\epsilon_{d,k} = \mathbb{E}[\vert J_{d,k}^{\text{mMTC+}} \vert^2]$, $\varepsilon_{d,u} = \mathbb{E}[\vert J_{d,u}^{\text{eMBB+}} [n]\vert^2]$, $\chi_d = \mathbb{E}[\vert W_d^{\text{mMTC+}} \vert^2]$ are the strength of the (mMTC+/eMBB+) interfering signals and thermal noise, respectively. Unlike \eqref{eq:14}, now the focus lies on the so-called spreading gain $N$. A more comprehensive discussion is provided in the simulations.

In contrast to the previous case, evaluating high-order moments for mMTC+ devices can be cumbersome; for example, expectations involving four inner/outer products (i.e., eighth-order moments) \cite{Mal11}. While these terms can be readily handled through numerical methods (by approximating statistical expectations with sample averages \cite{Ozd19}), closed-form expressions can be obtained when specializing to uncorrelated fading, leading to the following results \cite[Section~III]{Lie25a}:
\begin{equation}
    \lambda_d = N^2L^2\left\vert \sum\nolimits_m b_{m,d} \hat{\beta}_{m,d}^2 \tilde{\beta}_{m,d,d} \left(\bar{\beta}_{m,d} + L \tilde{\beta}_{m,d,d} \right) \right\vert^2,            
    \label{eq:18}
\end{equation}
with $\hat{\beta}_{m,d} = \mathrm{tr}(\mathbf{B}_{m,d})/L$ for convenience and $\mathbf{B}_{m,d} \in \mathbb{C}^{L \times L}$ the MMSE matrix for device $d$ (cf. \eqref{eq:3}). Analogously, we also define $\bar{\beta}_{m,d} = \sum_k \tilde{\beta}_{m,d,k} + \sum_u \tilde{\alpha}_{m,d,u} + \sigma_m^2$, where $\tilde{\beta}_{m,d,k} = \zeta_k \beta_{m,k} \vert \bm{\pi}_k^{\mathrm{H}} \bm{\pi}_d\vert^2$ and $\tilde{\alpha}_{m,d,u} = \eta_u \alpha_{m,u} \vert \bm{\phi}_u^{\mathrm{H}} \bm{\pi}_d \vert^2$. The other terms in \eqref{eq:17} are given in \eqref{eq:19} at the top of page 8. 

It is worth noting that, while \eqref{eq:15} generalizes the conventional eMBB+ framework by accounting for mMTC+ interference, to the best of our knowledge, \eqref{eq:18} and \eqref{eq:19} are novel within the CF-mMIMO literature. This, in turn, paves the way for potential advances in the coexistence of 6G technologies. Finally, as mentioned earlier, the data rate of device $d$ under the FBL regime (in bits/Hz) can be approximated by \cite{Pol10}
\setcounter{equation}{19}
\begin{equation}
    R_d^{\text{mMTC+}} (\rho_d) \approx C(\rho_d) - v_d D(\rho_d),
    \label{eq:20}
\end{equation}
where $C(x) \triangleq \log_2 (1 + x)$ is the channel capacity; $D(x) \triangleq \sqrt{V(x)}$ is the FBL penalty, with $V(x) \triangleq 2 x(1 + x)^{-1}$ the channel dispersion \cite{Sca17}; and $v_d \triangleq (\log_2 e / \sqrt{n_d}) Q^{-1}(\text{PER}_d)$ is a scaling (constant) factor to ease notation, with $Q(\cdot)$ the Gaussian Q-function, $\text{PER}_d$ the packet error rate (or decoding error probability), and $n_d$ the number of transmit symbols carrying actual information (i.e., $z_d$).

\section{Problem Formulation} \label{sec:4}
The purpose of this paper is to maximize the minimum EE of the mMTC+ devices subject to four constraints: $C1$ and $C2$ limit the UL transmit powers, $C3$ and $C4$ ensure all terminals have rates above a certain QoS threshold, whereas $C5$ guarantees a minimum SINR for reliable mMTC+ packet decoding (otherwise, too many decoding errors might occur, leading to possible communication failure). By denoting $\bm{\theta}$ as the vector containing the coefficients $p_u$ and $q_d$, the eMBB+ and mMTC+ data rates can be written as $R_u^{\text{eMBB+}}(\bm{\theta}) \equiv R_u^{\text{eMBB+}}(\gamma_u(\bm{\theta}))$ and $R_d^{\text{mMTC+}}(\bm{\theta}) \equiv R_d^{\text{mMTC+}}(\rho_d(\bm{\theta}))$, respectively. Accordingly, the optimization can be formulated as follows:
\begin{align}    
    \underset{\bm{\theta}}{\max} \, \underset{d}{\min} \, & \frac{\psi}{N} \frac{R_d^{\text{mMTC+}}(\bm{\theta})}{\mu_d q_d + \Theta_d} & \nonumber \\
    \text{s.t.} \quad &C1: 0 \leq q_d \leq Q_d,  \, \forall d, \quad C2: 0 \leq p_u \leq P_u, \, \forall u \nonumber \\        
    \label{eq:21} &C3: \left(\psi/N\right) R_d^{\text{mMTC+}}(\bm{\theta}) \geq R^{\text{mMTC+}}, \, \forall d, \\ 
    &C4: \psi R_u^{\text{eMBB+}}(\bm{\theta}) \geq R^{\text{eMBB+}}, \, \forall u, \nonumber \\ 
    &C5: \rho_d \geq S, \, \forall d, \nonumber   
\end{align}
with $\psi \triangleq B\tau_u/\tau_c$ the factor compressing the number of (UL) transmit symbols $\tau_u$, the number of coherence samples $\tau_c$ (per PRB), and the system's bandwidth $B$. Note that the factor $1/N$ represents the cost of the proposed spreading, i.e., using the PN sequences requires $N$-fold more (time/frequency) samples \cite{Jin08}. Therefore, unlike SINRs, the “penalized” mMTC+ rates decrease with increasing $N$. As discussed later, this introduces a relevant trade-off when choosing the optimal $N$: short sequences compromise the spreading gain (low SINR) but have little to no effect on the data rate (a small penalty). Besides, $\mu_d$ represents the inefficiency of the power amplifier and $\Theta_d$ the total static power consumption of terminal $d$ \cite{Mat20}.

As mentioned earlier, the above problem will be initially addressed using model-based approaches and subsequently using data-driven algorithms. This is discussed in the following.

\section{Analytical Solution} \label{sec:5}
In our previous work \cite{Lie25a}, the mMTC+ performance was characterized by Shannon's capacity. This allowed us to find a global optimum using a bisection search and to solve a sequence of linear feasibility problems \cite[Algorithm~4.1]{Boy04}. 

In contrast, some challenges arise in this setting. First, the objective is no longer the minimum data rate but the minimum EE. These types of problems involve multiple ratios of functions and are commonly known as \textit{generalized} fractional programs \cite{Cro91}. As we will see, the optimization can be solved globally in polynomial time only under special circumstances, using Dinkelbach's algorithm \cite{Din67}. 

\pagebreak

More precisely, when dealing with a problem of the form
\begin{equation}    
    \underset{\mathbf{x}}{\max} \, \underset{k = 1,\ldots,K}{\min} \, \frac{f_k(\mathbf{x})}{g_k(\mathbf{x})} \quad \text{s.t.} \quad \mathbf{x} \in \mathcal{D},
    \label{eq:22}
\end{equation}
where $\mathbf{x}$ is the design variable and $\mathcal{D} \subseteq \mathbb{R}^N$ is the constraint set such that $f_k: \mathcal{D} \to \mathbb{R}$ and $g_k: \mathcal{D} \to \mathbb{R}_{++}$ $\forall k = 1,\ldots,K$, the solution exists and is given by \cite[Proposition 1]{Zap17}
\begin{equation}
    \mathbf{x}^\star = \underset{\mathbf{x} \in \mathcal{D}}{\mathrm{argmax}} \, \underset{k=1,\ldots,K}{\min} \, f_k(\mathbf{x}) - \lambda^\star g_k(\mathbf{x}),
    \label{eq:23}
\end{equation}
with $\lambda^\star$ the unique zero of the auxiliary function
\begin{equation}
    F(\lambda) = \underset{\mathbf{x} \in \mathcal{D}}{\max} \, \underset{k = 1,\ldots,K}{\min} \, f_k(\mathbf{x}) - \lambda g_k(\mathbf{x}).
    \label{eq:24}
\end{equation}

\begin{figure*}[t]
    \begin{align}    
        \hat{z}_d &= \sqrt{q_d} \underbrace{\mathbb{E}\left[\sum_{n,m} b_{m,d} \left\vert \mathbf{t}_{m,d}^{\mathrm{H}}[n] \mathbf{g}_{m,d}[n] \right\vert^2 \right]}_{\triangleq S_d}z_d + \sqrt{q_d} \underbrace{\left( \sum_{n,m} b_{m,d} \left\vert \mathbf{t}_{m,d}^{\mathrm{H}}[n] \mathbf{g}_{m,d}[n] \right\vert^2 - S_d\right)}_{\triangleq V_d} z_d + \underbrace{\sum_{n,m} c_d^*[n] b_{m,d} \tilde{\mathbf{t}}_{m,d}^{\mathrm{H}}[n] \mathbf{w}_m[n]}_{\triangleq W_d^{\text{mMTC+}}} \nonumber \\      
        &\quad + \sum_{k \neq d} \sqrt{q_k} \underbrace{\sum_n c_d^*[n] c_k[n] \sum_m b_{m,d} \tilde{\mathbf{t}}_{m,d}^{\mathrm{H}}[n] \mathbf{g}_{m,k}[n]}_{\triangleq J_{d,k}^{\text{mMTC+}}} z_k  + \sum_u \sqrt{p_u} \sum_n \underbrace{c_d^*[n] \sum_m b_{m,d} \tilde{\mathbf{t}}_{m,d}^{\mathrm{H}}[n] \mathbf{h}_{m,u}[n]}_{\triangleq J_{d,u}^{\text{eMBB+}} [n]} s_u[n],     
        \label{eq:16} \tag{16}
    \end{align}
\vspace{-2mm}
\hrule
\end{figure*}

This means problem \eqref{eq:22} can be globally solved by finding the unique zero of $F(\lambda)$. For that task, a popular option is the generalized Dinkelbach's algorithm, reported in Algorithm~\ref{alg:1}, whose convergence is guaranteed provided that the individual problems in \eqref{eq:23} can be solved globally. If $f_k(\mathbf{x})$ is concave, $g_k(\mathbf{x})$ is convex, and all constraints are also convex (that is, if the optimization in \eqref{eq:23} is convex), this can be accomplished in polynomial time and linear convergence rate \cite{Cro91}.

\begin{algorithm}[t]
\begin{algorithmic}[1]       
    \State Choose convergence tolerance threshold $F_0 > 0$
    \State Initialize $\lambda$ with $F(\lambda)\geq 0$
    \While{$F(\lambda) > F_0$}
        \State Solve \eqref{eq:23} to find $\mathbf{x}^\star$ with $\lambda^\star = \lambda$
        \State Update $\lambda$ as $\min_k \, f_k(\mathbf{x}^\star)/g_k(\mathbf{x}^\star)$
    \EndWhile
\end{algorithmic}
\caption{Generalized Dinkelbach's algorithm \cite{Din67}}
\label{alg:1}
\end{algorithm}

Unfortunately, the conditions above cannot be met even in the simple case where the packets have infinite lengths. In a nutshell, the functions $f_k(\mathbf{x}) \equiv C(\rho_d)$ are not concave w.r.t. optimization variables $\mathbf{x} \equiv \bm{\theta}$. Moreover, in the FBL regime, the additional penalty $D(\rho_d)$ in \eqref{eq:20} is not convex either. 

To confront this, one could directly use standard global optimization techniques (or general-purpose programming solvers) to find the global optimum. However, in addition to their prohibitive complexity, convergence is guaranteed only if the functions exhibit limited variability. Therefore, we will resort to suboptimal methods to obtain a feasible solution. That is, using a series of convex/concave bounds, we iteratively update our solution until convergence to a stationary point. At each step, we will be applying FP tools, a procedure commonly known as \textit{sequential} FP.

For clarity of the explanation, we begin by considering the case $n_d \to \infty$ and then extend the analysis to the FBL regime. This will be discussed in the upcoming subsections.

\subsection{Infinite blocklength regime} \label{sec:5.1}
When $n_d \to \infty$, we can write the maximin optimization as
\begin{equation}    
    \underset{\bm{\theta}}{\max} \, \underset{d}{\min} \, \frac{\psi}{N} \frac{\log_2\left(1 + \rho_d(\bm{\theta})\right)}{\mu_d q_d + \Theta_d} \quad \text{s.t.} \quad \bm{\theta} \in \mathcal{C}
    \label{eq:25}
\end{equation}
where $\mathcal{C}$ is the feasible set defined by constraints $C1-C5$. 

As mentioned before, the numerator in the objective above is not concave w.r.t. $\bm{\theta}$. To circumvent this issue, we will derive a concave lower bound based on the following inequality \cite{Lie25b}
\begin{equation*}
    \ln \left(1 + \frac{x}{y}\right) \geq \ln \left(1 + \frac{\bar{x}}{\bar{y}}\right) + \frac{\bar{x}}{\bar{y}} \left(2\sqrt{\frac{x}{\bar{x}}} - \frac{x + y}{\bar{x} + \bar{y}} - 1\right), \\    
\end{equation*}
for any $x > 0$, $y > 0$, $\bar{y} > 0$, and $\bar{x} > 0$. 

\begin{figure*}[t]
\begin{align}
    \nu_d &=  N L \sum\nolimits_m b_{m,d} \hat{\beta}_{m,d}^4 \tilde{\beta}_{m,d,d}^2 \left((L + 1)\left((L + 1)\tilde{\beta}_{m,d,d}\left(L\tilde{\beta}_{m,d,d} + 4\bar{\beta}_{m,d}\right) + 2\bar{\beta}_{m,d}^2\right) - L\left(\bar{\beta}_{m,d} + L \tilde{\beta}_{m,d,d}\right)^2\right), \nonumber \\
    \epsilon_{d,k} &=  N L \zeta_d \sum\nolimits_m b_{m,d} \hat{\beta}_{m,d}^4 \tilde{\beta}_{m,d,d} \beta_{m,k} \left((L + 1)\bar{\beta}_{m,d}^2 + (L + 1)^2\bar{\beta}_{m,d}\left(\tilde{\beta}_{m,d,d} + \tilde{\beta}_{m,d,k}\right) + L\left(2L + 1\right) \tilde{\beta}_{m,d,d} \tilde{\beta}_{m,d,k}\right) \nonumber \\
    &\quad + L^4\zeta_d \zeta_k \left\vert \bm{\pi}_k^{\mathrm{H}} \bm{\pi}_d \right\vert^2 \left\vert \sum\nolimits_m b_{m,d} \hat{\beta}_{m,d}^2 \tilde{\beta}_{m,d,d} \beta_{m,k} \right\vert^2, \nonumber \\
    \varepsilon_{d,u} &= NL \zeta_d \sum\nolimits_m b_{m,d} \hat{\beta}_{m,d}^4 \tilde{\beta}_{m,d,d} \alpha_{m,u} \left((L + 1)\bar{\beta}_{m,d}^2 + (L + 1)^2\bar{\beta}_{m,d}\left(\tilde{\beta}_{m,d,d} + \tilde{\alpha}_{m,d,u}\right) + L\left(2L + 1\right) \tilde{\beta}_{m,d,d} \tilde{\alpha}_{m,d,u}\right) \nonumber \\
    &\quad + NL^4\zeta_d \eta_u \left\vert \bm{\phi}_u^{\mathrm{H}} \bm{\pi}_d \right\vert^2 \left\vert \sum\nolimits_m b_{m,d} \hat{\beta}_{m,d}^2 \tilde{\beta}_{m,d,d} \alpha_{m,u} \right\vert^2, \nonumber \\
    \chi_d &= NL(L + 1) \zeta_d \sum\nolimits_m b_{m,d} \hat{\beta}_{m,d}^4 \tilde{\beta}_{m,d,d} \bar{\beta}_{m,d} \sigma_m^2\left((L + 1)\tilde{\beta}_{m,d,d} + \bar{\beta}_{m,d}\right),
    \label{eq:19} \tag{19} 
\end{align}
\vspace{-2mm}
\hrule
\vspace{-2mm}
\end{figure*}

Let us also rewrite the SINR as $\rho_d(\bm{\theta}) = \lambda_d q_d/\varrho_d(\bm{\theta})$ with
\begin{equation}
    \varrho_d(\bm{\theta}) \triangleq \nu_d q_d + \sum\nolimits_{k \neq d} \epsilon_{d,k} q_k + \sum\nolimits_u \varepsilon_{d,u} p_u + \chi_d.
    \label{eq:26}
\end{equation}

At each iteration, the set of power coefficients $\bm{\theta}$ is gradually updated from the previous feasible point $\bar{\bm{\theta}}$. In that sense, when applying the previous bounds to the capacity for $x=\lambda_d q_d$, $y=\varrho_d(\bm{\theta})$, $\bar{x}=\lambda_d \bar{q}_d$, and $\bar{y}=\varrho_d(\bar{\bm{\theta}})$, one can then write
\begin{equation}    
    C_d(\bm{\theta}) \geq \log_2\left(1 + \frac{\displaystyle \lambda_d \bar{q}_d}{\displaystyle \varrho_d(\bar{\bm{\theta}})}\right) + \frac{1}{\displaystyle\ln 2}\frac{\displaystyle \lambda_d \bar{q}_d}{\displaystyle \varrho_d(\bar{\bm{\theta}})} \left(\Gamma(\bm{\theta},\bar{\bm{\theta}}) - 1\right),
    \label{eq:27}
\end{equation}
where $C_d(\bm{\theta}) \equiv C(\rho_d(\bm{\theta})) = \log_2(1 + \lambda_d q_d/\varrho_d(\bm{\theta}))$ and 
\begin{equation}
    \Gamma(\bm{\theta},\bar{\bm{\theta}}) \triangleq 2\sqrt{\frac{\lambda_d q_d}{\lambda_d \bar{q}_d}} - \frac{\lambda_d q_d + \varrho_d(\bm{\theta})}{\lambda_d \bar{q}_d + \varrho_d(\bar{\bm{\theta}})}.
    \label{eq:28}
\end{equation}

It is easy to prove that the function $\Gamma(\bm{\theta},\bar{\bm{\theta}})$ is concave w.r.t. $\bm{\theta}$ (square root plus linear) and so does the lower bound in \eqref{eq:27}, say $\tilde{C}_d(\bm{\theta},\bar{\bm{\theta}})$. Hence, by replacing the objective in \eqref{eq:25} with this surrogate, we have the approximate optimization
\begin{equation}    
    \underset{\bm{\theta}}{\max} \, \underset{d}{\min} \, \frac{\psi}{N} \frac{\tilde{C}_d(\bm{\theta},\bar{\bm{\theta}})}{\mu_d q_d + \Theta_d} \quad \text{s.t.} \quad \bm{\theta} \in \mathcal{C},
    \label{eq:29}
\end{equation}
which can be cast as a (maximin) concave fractional problem \cite{Avr03}: maximization of a pseudo-concave objective function (ratio of concave and affine) subject to convex constraints.

\begin{proof} 
Constraints $C1$, $C2$, and $C5$ are convex by definition (bounds on linear functions). A similar argument applies to $C3$ and $C4$ due to the monotonically increasing nature of the logarithm. In particular, we can write both constraints as
\begin{equation}
    \begin{aligned}
        &C3: \rho_d(\bm{\theta}) \geq 2^{\left(N/\psi\right)R^{\text{mMTC+}}} - 1, \quad &\forall d, \\ 
        &C4: \gamma_u(\bm{\theta}) \geq 2^{\left(1/\psi\right)R^{\text{eMBB+}}} - 1, \quad &\forall u,
    \end{aligned}
    \label{eq:30}
\end{equation}
which are also convex (SINRs are linear ratios).
\end{proof}

Among other mild assumptions, by sequentially updating $\bm{\theta}$ with the global optima of \eqref{eq:29}, we will converge to a stationary point of \eqref{eq:25} \cite[Section~IV]{Mat20}. At this point, we can thus safely apply Dinkelbach's transform as per \eqref{eq:23}, i.e.,
\begin{equation}
    \bm{\theta}^\star = \underset{\bm{\theta} \in \mathcal{C}}{\mathrm{argmax}} \, \underset{d}{\min} \, \left(\psi/N\right) \tilde{C}_d(\bm{\theta},\bar{\bm{\theta}}) - \theta^\star\left(\mu_d q_d + \Theta_d\right),
    \label{eq:31}
\end{equation}
where $\theta^\star$ is the zero of the auxiliary function
\begin{equation}
    F(\vartheta) = \underset{\bm{\theta}}{\max} \, \underset{d}{\min} \, \left(\psi/N\right) \tilde{C}_d(\bm{\theta},\bar{\bm{\theta}}) - \vartheta\left(\mu_d q_d + \Theta_d\right).
    \label{eq:32}
\end{equation}

Overall, for every $\bar{\theta}$ in the sequential FP procedure (outer layer), we will use Dinkelbach's algorithm to find the unique root $\vartheta^\star$ (inner layer). Each step will then focus on the individual problems in \eqref{eq:31}, whose solution can be found via numerical methods, e.g., \cite{CVX24}. As summarized in Algorithm~\ref{alg:2}, we end with a series of loops that converge to a global optimum of the corresponding instance of the fractional problem in \eqref{eq:29}, which in turn will be a local solution of the original formulation \eqref{eq:25}. In line with the CF goal, this power control scheme will contribute to: (a) mMTC+ networks becoming more sustainable and battery-aware, plus (b) terminals having uniformly good QoS over the entire coverage area.

\begin{algorithm}[t]
\begin{algorithmic}[1]           
    \State Initialize power coefficients $\bm{\theta}, \bar{\bm{\theta}} \in \mathcal{C}$ and $\vartheta$ with $F(\vartheta) \geq 0$    
    \State Choose convergence tolerance thresholds $\theta_0, F_0 > 0$
    \While{$\| \bm{\theta} - \bar{\bm{\theta}} \|^2/ \| \bm{\theta} \|^2 > \theta_0$}
        \While{$F(\vartheta) > F_0$}            
            \State Solve \eqref{eq:31} with $\vartheta^\star = \vartheta$ and $\bar{\bm{\theta}}$ to find $\bm{\theta}$
            \State Update $\vartheta = \min_d \, (\psi/N)\tilde{C}_d(\bm{\theta},\bar{\bm{\theta}})/(\mu_d \bar{q}_d + \Theta_d)$
        \EndWhile
        \State Update $\bar{\bm{\theta}} = \bm{\theta}$ (optimal solution $\bm{\theta}^\star$)
    \EndWhile
\end{algorithmic}
\caption{Proposed sequential FP power control ($n_d \to \infty$)}
\label{alg:2}
\end{algorithm}

\subsection{Finite blocklength regime} \label{sec:5.2}
For the extension to short-packet transmissions, we will require additional approximations to address the FBL penalty. Provided the concave lower bound $\tilde{C}_d(\bm{\theta},\bar{\bm{\theta}})$ for the capacity, we will now need a convex upper bound for the term $D_d(\bm{\theta}) \equiv D(\rho_d(\bm{\theta})) = \sqrt{2 \rho_d(\bm{\theta})/(1 + \rho_d(\bm{\theta}))}$.

Inspired by the arithmetic-geometric mean inequality \cite[(28)]{Sun17}, we can write the following surrogate:
\begin{equation*}
    \prod_{i=1}^n x_i^{\alpha_i} \leq \left(\prod_{i=1}^n \bar{x}_i^{\alpha_i}\right) \sum_{i=1}^n \frac{\vert \alpha_i \vert}{\| \bm{\alpha} \|_1} \left(\frac{x_i}{\bar{x}_i}\right)^{\| \bm{\alpha} \|_1 \mathrm{sgn}(\alpha_i)},    
\end{equation*}
for $x_i > 0$, $\bar{x}_i > 0$, and $\alpha_i > 0$, $\forall i = 1,\ldots,n$. Note that $\bm{\alpha} = [\alpha_1,\ldots,\alpha_n]$ is the concatenation of exponents.

\pagebreak

Particularizing for $n = 2$, $x_1 = \lambda_d q_d$, $x_2 = \lambda_d q_d + \varrho(\bm{\theta})$, $\bar{x}_1 = \lambda_d \bar{q}_d$, $\bar{x}_2 = \lambda_d \bar{q}_d + \varrho(\bar{\bm{\theta}})$, and $\bm{\alpha} = [0.5, -0.5]$, we have
\begin{equation}
    D_d(\bm{\theta}) \leq \frac{\displaystyle D_d(\bar{\bm{\theta}})}{2} \left(\frac{\lambda_d \bar{q}_d + \varrho(\bar{\bm{\theta}})}{\lambda_d q_d + \varrho(\bm{\theta})}+\frac{\lambda_d q_d}{\lambda_d \bar{q}_d}\right) \triangleq \tilde{D}_d(\bm{\theta},\bar{\bm{\theta}}),
    \label{eq:33}
\end{equation}
which is convex w.r.t. $\bm{\theta}$ (convex-plus-linear) and also satisfies the conditions in \cite[Section~IV]{Mat20}. Together with the bound in \eqref{eq:27}, we can then formulate the approximate maximin as
\begin{equation}    
    \underset{\bm{\theta} \in \mathcal{C}}{\max} \, \underset{d}{\min} \, (\psi/N) (\tilde{C}_d(\bm{\theta},\bar{\bm{\theta}}) - v_d \tilde{D}_d(\bm{\theta},\bar{\bm{\theta}}))/(\mu_d q_d + \Theta_d),
    \label{eq:34}
\end{equation}
which, again, can be shown to be a concave fractional problem. However, unlike before, constraint $C3$ is also convexified, i.e.,
\begin{equation}
    C3: \tilde{C}_d(\bm{\theta},\bar{\bm{\theta}}) - v_d \tilde{D}_d(\bm{\theta},\bar{\bm{\theta}}) \geq \left(N/\psi\right) R^{\text{mMTC+}}, \quad \forall d.
    \label{eq:35}
\end{equation}

Finally, by following the steps outlined in the previous subsection, we can find a first-order optimal solution to the original problem. Briefly, we can apply the steps in Algorithm~\ref{alg:2}: sequential FP on top of Dinkelbach's method.

\section{GNN Implementation} \label{sec:6}
Although the previous analytical solution yields a KKT point, its complexity becomes impractical as $M$, $K_u$, and $K_d$ increase. This fact poses a serious challenge for the deployment of real-time future networks. 

As in \cite{Mis24}, this work aims to leverage a GNN to learn an efficient approximation of the analytical solution. Specifically, the GNN will be designed to achieve EE performance comparable to that of the sequential FP, while dramatically reducing the computational burden, even though the resulting power coefficients need not be identical. The power control policy will also be designed solely based on the LSF\footnote{As per the existing literature, the analysis is conducted under a moderate mobility regime, where the LSF is assumed to vary slowly and can therefore be considered constant over consecutive time intervals \cite{Nad23}.} vector $\bm{\varphi}$ (the concatenation of the coefficients $\alpha_{m,u}$ and $\beta_{m,d}$). Hence, both the sequential FP baseline and the proposed GNN approach impose the same requirements on the fronthaul\footnote{A detailed discussion of fronthaul load and synchronization aspects can be found in \cite{Dem21}. In the considered framework, the impact of these constraints is identical for both the GNN-based and FP-based solutions, since they share the same optimization formulation and operate on the same set of LSF inputs. Moreover, thanks to the terminal-centric architecture, the signaling exchange remains limited, regardless of the network's size (cf. \cite[Table~I]{Bjo20a}).}.

In what follows, we describe the graph representation of the CF-mMIMO system, the structure of the proposed GNN, and the training procedure.

\subsection{Graph Representation} \label{sec:6.1}
We generate datasets of tuples ($\bm{\varphi}$, $\bm{\theta}$) drawn from distinct simulation environments. Interestingly, any permutation applied to the LSF coefficients is also applied to power coefficients. Thus, the underlying learning task does not depend on any specific ordering of the inputs, a property known as \textit{permutation equivariance}. Since GNNs inherently respect this symmetry, they are particularly well-suited to address power control problems \cite{Mis24}.

\begin{figure}[t]
    \centerline{\includegraphics[scale=0.5, trim=0.75cm 0 0 0.25cm, clip]{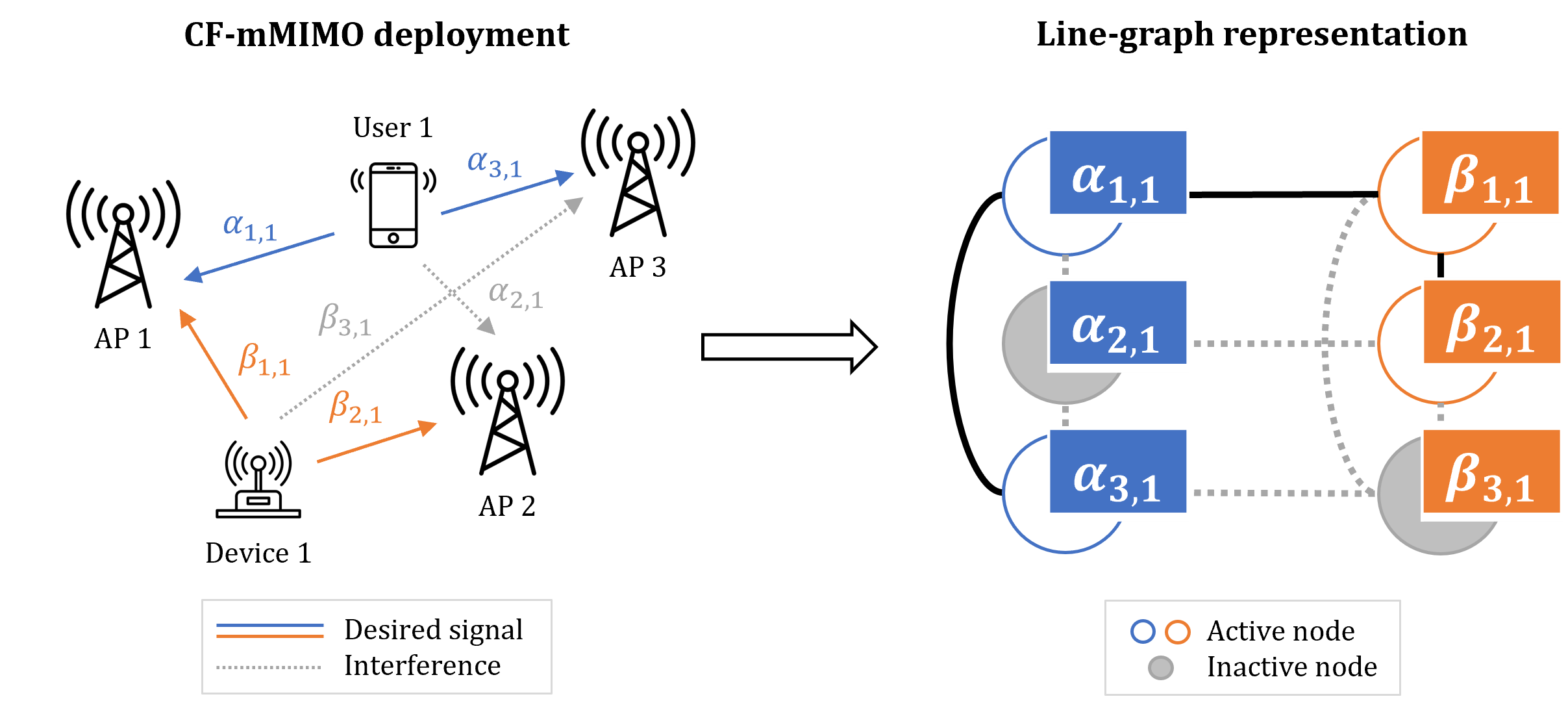}} 
    \caption{Illustrative example of the line-graph representation of a terminal-centric CF-mMIMO deployment where $K_u = 1$ eMBB+ user and $K_d = 1$ mMTC+ device are simultaneously served by $M = 3$ single-antenna APs.}
    \label{fig:2}
    \vspace{-2mm}
\end{figure}

Training GNNs requires expressing both inputs and outputs in graph-based form. Rather than modeling the direct bipartite connections between APs and terminals, we adopt a line-graph representation in which each link is mapped to a node. An example is depicted in Fig. \ref{fig:2}. This transformation provides a richer description of the power control problem by explicitly reflecting the channel interactions among related terminals and APs. Accordingly, the data instances are converted into \textit{heterogeneous directed} graphs $\mathcal{G} = (\mathcal{V}_u, \mathcal{V}_d, \mathcal{E})$, where $\mathcal{V}_u$ denotes the set of AP-user nodes, $\mathcal{V}_d$ is the set of AP-device nodes, and $\mathcal{E}$ represents the set of directed edges.

Note that, unlike \cite{Mis24}, here we have two (disjoint) subsets of nodes: $\mathcal{V}_u = \{1,\ldots,MK_u\}$ for the AP-user pairs $(m,k_u) \in \{1,\ldots,M\}\times\{1,\ldots,K_u\}$, and $\mathcal{V}_d = \{1,\ldots,MK_d\}$ for the AP-device pairs $(m,k_d) \in \{1,\ldots,M\}\times\{1,\ldots,K_d\}$. The one-to-one mappings between these pairs and sets are given by $\varpi_u(m,k_u)=i$ and $\varpi_d(m,k_d)=j$, respectively. In that sense, nodes $i\in \mathcal{V}_u, j \in \mathcal{V}_d$ are associated with tensors $h_i(t),g_j(t)$ called \textit{node features}, which are successively updated through $T$ iterations. For instance, at $t = 0$, we have $h_i(0) = \alpha_{m,k_u}$ and $g_j(0) = \beta_{m,k_d}$ (the LSF inputs of the problem). 

Let us define $\mathbf{h}(t)$ and $\mathbf{g}(t)$ as the tensors containing all the node features $h_i(t), i\in \mathcal{V}_u$ and $g_j(t), j\in \mathcal{V}_d$ at iteration $t$, respectively. Then, after the intermediate steps $t = 1,\ldots,T-1$, known as \textit{hidden features} or \textit{hidden layers}, the goal is to approximate the outputs $\bm{\theta}$ with the final features, i.e., 
\begin{equation}
    \hat{\bm{\theta}} \triangleq [\Upsilon_u(\mathbf{h}(T)), \Upsilon_d(\mathbf{g}(T))] \approx \bm{\theta},
    \label{eq:36}
\end{equation}
where the readout functions $\Upsilon_u(\cdot)$ and $\Upsilon_d(\cdot)$ transform the $MK_u$ AP-user and $MK_d$ AP-device features into the power coefficients $\{p_u\} \in \mathbb{R}^{K_u}$ and $\{q_d\} \in \mathbb{R}^{K_d}$, respectively.

Since the GNN performs local computations on each node and its neighbors, we connect two nodes if they share the same AP, user, or device (without self-loops). That is, we distinguish between six types of edges $e \in \mathcal{E}$:
\begin{equation*}
    \text{type}(e) = \left\{
        \begin{array}{ll}
            \text{AP}_{u,u}, & e=(\varpi_u(m,k_u),\varpi_u(m,k_u')) \\
            \text{AP}_{d,d}, & e=(\varpi_d(m,k_d),\varpi_d(m,k_d')) \\
            \text{AP}_{u,d}, & e=(\varpi_u(m,k_u),\varpi_d(m,k_d)) \\
            \text{AP}_{d,u}, & e=(\varpi_d(m,k_d),\varpi_u(m,k_u)) \\
            \text{user}, & e=(\varpi_u(m,k_u),\varpi_u(m',k_u)) \\
            \text{device}, & e=(\varpi_d(m,k_d),\varpi_d(m',k_d)).
        \end{array}
    \right.
\end{equation*}

Let $\mathcal{T} \triangleq \{\text{AP}_{u,u}, \text{AP}_{d,d}, \text{AP}_{u,d}, \text{AP}_{d,u}, \text{user}, \text{device}\}$ be the set of edge types, and $\mathcal{T}_u, \mathcal{T}_d \subset \mathcal{T}$ the (disjoint) subsets where the destination node has type AP-user, AP-device, respectively. Each type $\bullet \in \mathcal{T}$ will also determine the (disjoint) neighbor sets $\mathcal{N}_{\bullet} (i), \mathcal{N}_{\bullet} (j)$ of nodes $i \in \mathcal{V}_u, j \in \mathcal{V}_d$. As an example, $\mathcal{N}_{\text{user}}(i) = \{i' \in \mathcal{V}_u: (i,i') \in \mathcal{E}, \, \text{type}(i,i') = \text{user}\}$. We omit the remaining definitions to avoid redundancy.

With this construction, nodes with first-order dependencies are connected as neighbors, whereas nodes that share the same AP, user, or device are treated separately. To enable scalability, however, it is necessary to account for the fact that each AP typically serves only a subset of terminals. This requirement introduces an additional layer in the graph, since edges must now reflect not only their $\text{type}(e)$ but also whether the corresponding AP-terminal link is active or inactive.

To capture the activity of the source and destination nodes, we introduce \textit{edge attributes} $\text{attr}(e)$. This way, we can embed the associations through one-hot encodings, as summarized in Table~\ref{tab:1}. This representation, which is valid for both node types, enables the trained model to be more versatile and generalizable than other methods, such as masking (cf. \cite{Mis24}).

\begin{table} [t]
    \caption{One-hot encoding for the edge attributes.}
    \centering
    \begin{tabular}{|c|c|c|c|}
        \hline
        Source node $i$ & Destination node $i'$ & $\text{attr}(i,i')$\\
        \hline        
        Active & Active & 0001\\        
        Active & Inactive & 0010\\        
        Inactive & Active & 0100\\        
        Inactive & Inactive & 1000\\
        \hline
    \end{tabular}        
    \label{tab:1}    
\end{table}

\subsection{Network Structure} \label{sec:6.2}
\begin{figure}[t]    
    \centerline{\includegraphics[scale=1]{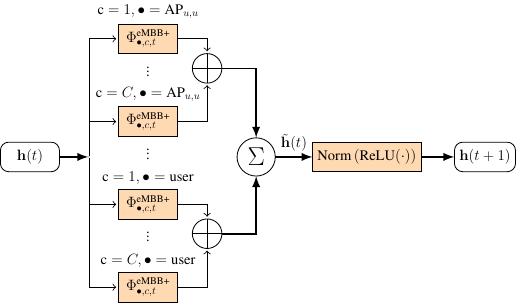}}
    \caption{Update of AP-user tensor $\mathbf{h}(t)$. A similar procedure holds for $\mathbf{g}(t)$.}
    \label{fig:3}
\end{figure}

In this paper, we adopt a graph transformer architecture similar to that of \cite{Mis24} and extend it to multiple node types. An overview is given in Fig.~\ref{fig:3}, where at iteration $t$, we uptade the two tensors $\mathbf{h}(t)$ and $\mathbf{g}(t)$ as
\begin{equation}
    \begin{aligned}
        \mathbf{h}(t + 1) &= \text{Norm}(\text{ReLU}(\tilde{\mathbf{h}}(t))), \\
        \mathbf{g}(t + 1) &= \text{Norm}(\text{ReLU}(\tilde{\mathbf{g}}(t))),
    \end{aligned}
    \label{eq:37}
\end{equation}
by applying ReLU activations followed by layer normalization to $\tilde{\mathbf{h}}(t), \tilde{\mathbf{g}}(t)$, the concatenations of the intermediate tensors that account for the multi-headed attention mechanism, i.e.,
\begin{equation}
    \begin{aligned}
        \tilde{h}_i(t) = \sum_{\bullet \in \mathcal{T}_u} \bigoplus_{c=1}^C \Phi_{\bullet,c,t}^{\text{eMBB+}}(i), \quad
        \tilde{g}_j(t) = \sum_{\bullet \in \mathcal{T}_d} \bigoplus_{c=1}^C \Phi_{\bullet,c,t}^{\text{mMTC+}}(j),
    \end{aligned}
    \label{eq:38}
\end{equation}
where $\bigoplus$ is the concatenation operator and $C \geq 1$ is the number of heads. Without loss of generality, we now omit the superscripts and focus on AP-user nodes (AP-device nodes can be described similarly). The aggregation function is then
\begin{equation}
    \Phi_{\bullet,c,t}(i) = \Lambda_{\bullet,c,t}^1(h_i(t)) +
    \sum\nolimits_{i'\in\mathcal{N}_{\bullet}(i)} \Psi_{\bullet,c,t}(i,i').
    \label{eq:39}
\end{equation}

In \eqref{eq:39}, $\Lambda$ denotes the linear operator and $\Psi_{\bullet,c,t}(i,i')$ is the message passing function between (AP-user) nodes $i, i' \in \mathcal{V}_u$: 
\begin{equation}
    \Psi_{\bullet,c,t}(i,i') = \alpha_{\bullet,c,t}(i,i') \left( \Lambda_{\bullet,c,t}^2(h_{i'}(t))+\Lambda_{\bullet,c,t}^3\left(\text{attr}(i,i')\right)\right),
    \label{eq:40}
\end{equation}
where $\alpha_{\bullet,c,t}(i,i')$ is the softmax attention coefficient given by \eqref{eq:41} at the top of the next page, with $\langle x, y \rangle = \exp(x^{\mathrm{T}} y/\sqrt{d})$ the exponential scale dot-product and $d$ the head size \cite{Vas17}. These coefficients regulate how neighboring features $h_{i'}(t)$ contribute to the update of $h_i(t+1)$, with values close to zero suppressing the feature. Note that the interaction between APs (or terminals) depends on factors such as spatial location and channel conditions, which is why we use attention with edge attributes to seize these heterogeneous relationships. In turn, multi-head attention enables different importance weights to be assigned to the features of the same neighbor.

\begin{figure*}[t]
\begin{equation}     
    \alpha_{\bullet,c,t}(i,i') = \frac{\langle
    \Lambda_{\bullet,c,t}^4(h_i(t)),
    (\Lambda_{\bullet,c,t}^5(h_{i'}(t))+ \Lambda_{\bullet,c,t}^3\left(\text{attr}(i,i') \right)
    \rangle}
    {\displaystyle \sum\nolimits_{i'\in\mathcal{N}_{\bullet,c,t}(i)}\langle
    \Lambda_{\bullet,c,t}^4(h_i(t)),
    (\Lambda_{\bullet,c,t}^5(h_{i'}(t))+ \Lambda_{\bullet,c,t}^3\left(\text{attr}(i,i') \right)
    \rangle},
    \label{eq:41}
\end{equation}
\vspace{-2mm}
\hrule
\end{figure*}

Finally, at step $T$, we apply the linear activation $h_i(T) = \Lambda_{\text{out}}(h_i(T-1))$, for all nodes $i\in V_u$. However, as mentioned earlier, we merge the predictions from the $MK_u$ tensors into the final $K_u$ outputs. This is accomplished with the transformation $\Upsilon_u$ from \eqref{eq:36}, which first averages over the $M$ APs and then applies a sigmoid function. In that sense, to comply with the power budget, we scale the outcome with $P_u$. As discussed in the next subsection, the other constraints can be addressed through the training loss.

Note that the choice of layer sizes and the number of heads is deployment-dependent and subject to hyperparameter tuning for each problem \cite{Cha20}. In our case, we found that $T = 7$ layers with sizes $[\text{in} = 1, 32, 32, 64, 64, 32, 32, \text{out} = 1]$ and $C = 4$  heads offered the best results and lowest complexity.

\subsection{Training Procedure} \label{sec:6.3}
The dataset is composed of $5$ scenarios\footnote{Due to the high complexity of the analytical solution, which is used to generate the groundtruth, we employ rather small deployments.} with tuples $K_u \in \{1,\ldots,5\}$, $K_d \in \{5,\ldots,10\}$, and $M \in \{5,\ldots,10\}$. Each setup comprises $10^4$ samples, divided into $80\%$, $10\%$, and $10\%$ for training, validation, and test, respectively. For training, we employ $l_2$ regularization with a weight decay of $10^{-4}$, and the AdamW optimizer \cite{Los19} with a learning rate of $10^{-4}$. The number of epochs is $100$, and the batch size is $32$.

In this work, we formulate a weighted training loss based on the mean squared errors (MSEs) w.r.t. the objective function and the logarithmic\footnote{Similar to \cite{Mis24}, we apply a $\log(\cdot)$ transformation to all input and output features. In addition, inputs are also normalized to standard scores.} power coefficients, i.e.,
\begin{equation}
   \mathcal{L}(\alpha) = \alpha\underbrace{\text{MSE}(\log(\hat{\bm{\theta}}),\log(\bm{\theta}))}_{\triangleq \mathcal{L}_{\bm{\theta}}}  + (1 - \alpha) \underbrace{\text{MSE}(\text{EE}(\hat{\bm{\theta}}),\text{EE}(\bm{\theta}))}_{\triangleq \mathcal{L}_{\text{EE}}}, 
   \label{eq:42}
\end{equation}
where $\alpha\in [0,1]$ is a tunable parameter, $\text{EE}(\cdot)$ is the energy efficiency, $\bm{\theta} \equiv \bm{\theta}^\star$ is the analytical solution, and $\hat{\bm{\theta}}$ is the GNN's prediction. Although the proposed function is continuous and differentiable, it does not guarantee satisfaction of the constraints in problem \eqref{eq:21}. More precisely, as shown by the simulations, using \eqref{eq:42} violates $C4$ (the QoS requirement on the users' data rate) $60\%$ of the time.

Because of the above, we modify the previous training loss using the augmented Lagrangian method \cite{Boy11}: 
\begin{equation}
    \mathcal{L}(\alpha,\beta) = \alpha \mathcal{L}_{\bm{\theta}} + (1 - \alpha) \mathcal{L}_{\text{EE}} + \beta \mathcal{L}_{\text{aug}},
    \label{eq:43}
\end{equation}
where $\beta\in \{0,1\}$ and $\mathcal{L}_{\text{aug}}$ is written as per (cf. \cite[(10.2)]{Eld22})
\begin{equation}
    \mathcal{L}_{\text{aug}} = \lambda v  + (\rho/2) v^2,
    \label{eq:44}
\end{equation}
with $\lambda$ the dual (Lagrangian) variable, $v = \textrm{ReLU}(R^{\text{eMBB+}} - \min_u R_u^{\text{eMBB+}}(\hat{\bm{\theta}}))$ the constraint violation, and $\rho$ the penalty parameter. Remarkably, note that $\lambda$ approximates the Lagrange multiplier and is updated with the projected gradient ascent at each training step, i.e., $\lambda \gets \max(0,\lambda + \rho v)$ \cite{Boy04}.

Thanks to the newly added loss $\mathcal{L}_{\text{aug}}$, the new function in \eqref{eq:43} will increase whenever the constraints are not met and, thus, the GNN prediction learns to satisfy the constraints while improving its EE (that is, $\hat{\bm{\theta}} \in \mathcal{C}$).

\section{Numerical Simulations} \label{sec:7}
In what follows, we present numerous experiments to validate the two main results: the MA scheme and power control optimization. In that sense, performance will be evaluated using the cumulative distribution functions (CDFs) of the mMTC+ EE and the eMBB+ data rate.

In all settings, terminals and APs are uniformly distributed within a $250$ m $\times$ $250$ m deployment area at fixed heights of $1.65$ m and $10$ m, respectively. We also assume that each terminal is associated with the $M_s$ APs with the largest LSF coefficients \cite{Elw23}. Other association rules can be found in \cite{DiG26}.

\begin{figure}[t]    
    \centerline{\includegraphics[scale=0.975]{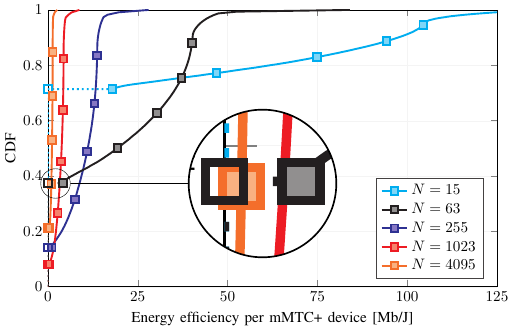}}
    \vspace{-2mm}
    \caption{CDF of the EE of mMTC+ devices vs. spreading factor $N$ under UPC. Dotted lines with empty markers ($\square$) indicate infeasible values.}    
    \label{fig:4}
    \vspace{-4mm}
\end{figure}

The scenario follows the micro-urban configuration from \cite{3GPP36814} with $P_u = 20$ dBm $\ll$ $Q_d = 10$ dBm $\forall u,d$, $\sigma_m^2 = N_o B$ $\forall m$, $N_o = -174$ dBm/Hz, and $B = 20$ MHz. The carrier frequency is $2$ GHz and the PN signatures are generated from \textit{m-sequences} with $N = 2^n - 1$, $n = 1,2,\ldots$ \cite[Section~5.4]{Sim94}. We consider PRBs of size $1$ ms and $200$ kHz \cite{Elw23}, i.e., $\tau_c = 200$ samples. The first $\tau_p = (K_u + K_d)/2$ symbols are used for channel estimation, while $\tau_u = (\tau_c - \tau_p)/2$ are allocated to UL communication. Besides the optimal power control (OPC) from Section~\ref{sec:5} and the GNN-based implementation\footnote{The associated source code and data files are available at: \url{https://github.com/Nokia-Bell-Labs/gnn-power-control-heterogeneous-cfmmimo}.} from Section~\ref{sec:6}, we include uniform power control (UPC) and fractional power control (FPC) as benchmark schemes \cite{Dem21}. We also extend the generalized FPC (G-FPC) from \cite{Lie26a} to the UL:
\begin{equation*}
    p_u = \frac{P_u}{c} \left(\sum_{m = 1}^M a_{u,m} \alpha_{u,m}\right)^\kappa, \, q_d = \frac{Q_u}{c} \left(\sum_{m = 1}^M b_{d,m} \beta_{d,m}\right)^\kappa,
\end{equation*}
where $c = \max_{u,d}((\sum_{m = 1}^M a_{u,m} \alpha_{u,m})^\kappa,(\sum_{m = 1}^M b_{d,m} \beta_{d,m})^\kappa)$ normalizes the power coefficients and $\kappa \in [-1,1]$ tunes the power distribution. To resemble a maximin operation, $\kappa < 0$ is preferable \cite[(7.34)]{Dem21}. For fairness, all benchmark schemes are modified to satisfy constraints $C1-C5$.

Unless otherwise stated, we fix $K_u = 2$, $K_d = 10$, $M = 10$, $M_s = 5$, $L = 4$, $R^{\text{eMBB+}} = 1$ Mbps, $R^{\text{mMTC+}} = 10$ kbps. Note that, given the sporadic nature of mMTC+ \cite{Hua22}, $K_d$ is the number of (active) devices transmitting simultaneously, which is significantly smaller than the total number of devices, say $D$. For instance, in NB-IoT applications with periodic reports of $l = 10$ kb every $t = 2$ h, $K_d = 10$ is equivalent to supporting $D = (R^{\text{mMTC+}}t/l) K_d > 70^3 $ terminals over the $250$ m $\times$ $250$ m deployment area (that is, $D > 10^6$ over $1$ km\textsuperscript{2}). We also consider $n_d = 100$ transmit symbols and a target PER of $10^{-3}$. Lastly, recall that 3GPP recommends low-order constellations for mMTC+, e.g., QPSK requires an SINR of around $S = 0$ dB to achieve a block error probability below $10\%$ \cite[Table~4.7]{Gho11}.

\subsection{Multiple Access Scheme} \label{sec:7.1}
To assess the performance of the MA scheme, we first analyze the impact of the spreading factor, shedding light on relevant trade-offs in system design. Later, to further emphasize the soundness and benefits of our spread-spectrum proposal, we will also include orthogonal counterparts with different physical resource allocations.

\begin{figure}[t]    
    \centerline{\includegraphics[scale=0.975]{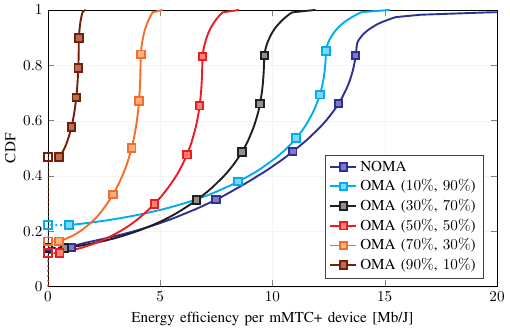}}
    \vspace{-2mm}
    \caption{CDF of the EE of mMTC+ devices vs. MA scheme (NOMA/OMA) under UPC. Dotted lines with empty markers ($\square$) indicate infeasible values.}    
    \label{fig:5}
\end{figure}

\begin{figure}[t]    
    \centerline{\includegraphics[scale=0.975]{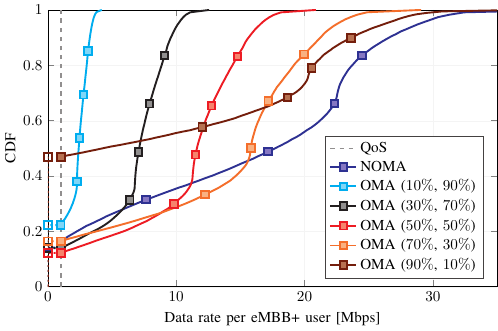}}
    \vspace{-2mm}
    \caption{CDF of the rate of eMBB+ users vs. MA scheme (NOMA/OMA) under UPC. Dotted lines with empty markers ($\square$) indicate infeasible values.}
    \label{fig:6}
    \vspace{-4mm}
\end{figure}

\subsubsection{Spreading Factor} Fig.~\ref{fig:5} illustrates the EE of the devices w.r.t. the number of PRBs $N$ (or, equivalently, the spreading factor) under UPC. Depending on the choice of $N$, the constraints might not be satisfied ($\text{EE} = 0$). For better readability, this is represented with dotted lines. As we can observe, a low $N$ leads to higher EE but also to a larger percentage of infeasible values (e.g., $N = 15$ yields more than $70\%$). This is because the SINR is too low for correct decoding ($C5$). In fact, the case of no spreading, i.e., $N = 1$, is not even shown since it entails no feasible points at all. On the other hand, for high numbers of PRBs, large transmit powers are also needed to compensate for the penalty coefficient $1/N$ in the pre-log term of the mMTC+ devices' rate. Consequently, a smaller EE is obtained. The figure thus reveals an optimal value of $N$ that maximizes the EE while reducing the number of infeasible values. In particular, the figure shows that, for the case at hand, $N=255$ achieves the best trade-off. Note that, as discussed later, considering other power control methods could further improve feasibility.

\subsubsection{Resource Allocation} Figs.~6 and 7 depict the mMTC+ EE and the eMBB+ rate w.r.t. the proposed spread-spectrum technique (non-orthogonal MA or NOMA) and the case where mMTC+ and eMBB+ use separate resources (orthogonal MA or OMA). For the latter, we assign $r_u$ ($\%$) and $r_d$ ($\%$) of the available PRBs to users and devices, respectively. In the plot, this is indicated by “OMA ($r_u$, $r_d$)”. Unsurprisingly, when fewer resources are allocated to mMTC+ devices, EE decreases, while eMBB+ rate improves (and vice versa). However, for extreme cases (e.g., $r_u = 90\%$ and $r_d = 10\%$), more values are infeasible. Remarkably, our approach outperforms the orthogonal counterpart in most cases, in both EE and rate, while maintaining (almost) the same feasibility. This result further strengthens the use of NOMA schemes (rather than OMA techniques such as network slicing) for the coexistence of heterogeneous services in 6G.

\begin{figure}[t]    
    \centerline{\includegraphics[scale=0.975]{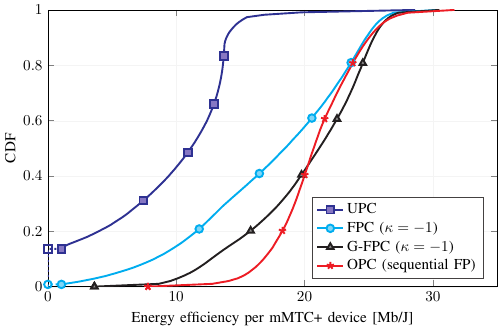}}
    \vspace{-2mm}
    \caption{CDF of the EE of mMTC+ devices vs. power control design. Dotted lines with empty markers ($\square$) indicate infeasible values.}    
    \label{fig:7}
\end{figure}

\begin{figure}[t]    
    \centerline{\includegraphics[scale=0.975]{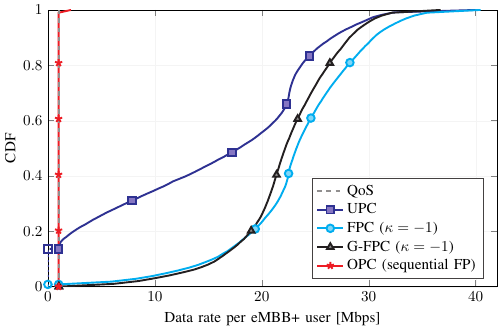}}
    \vspace{-2mm}
    \caption{CDF of the rate of eMBB+ users vs. power control design. Dotted lines with empty markers ($\square$) indicate infeasible values.}
    \label{fig:8}
    \vspace{-4mm}
\end{figure}

\subsection{Power Control Optimization} \label{sec:7.2}
In this subsection, we compare the performance of the model-based approach with that of the data-driven algorithm. 
\subsubsection{Sequential FP Solution} We start by analyzing the EE and data rate of the analytical OPC solution from Section~\ref{sec:5}. This is shown in Figs.~\ref{fig:7} and \ref{fig:8}, respectively. At a glance, we can readily see that our iterative methodology surpasses the (minimum) EE of all benchmarking schemes. More precisely, at least $40\%$ of the mMTC+ devices have better performance. This becomes more evident in the rate results, which show that the QoS constraint $C4$ is indeed achieved with equality, whereas the other policies unnecessarily exceed this threshold. In other words, thanks to our formulation, there is room to improve EE while guaranteeing minimum rates. Notably, feasibility is also enhanced relative to the UPC baseline (the other strategies entail few to no infeasible solutions). Lastly, the novel G-FPC has the highest EE among the heuristic schemes, so it is a valid alternative with negligible complexity.

\begin{figure}[t]    
    \centerline{\includegraphics[scale=0.975]{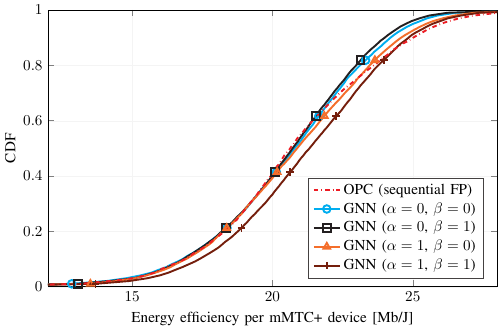}}
    \vspace{-2mm}
    \caption{CDF of the EE of mMTC+ devices vs. GNN training loss.}    
    \label{fig:9}
    \vspace{-4mm}
\end{figure}

\subsubsection{GNN Implementation} 
\begin{figure}[t]    
    \centerline{\includegraphics[scale=0.975]{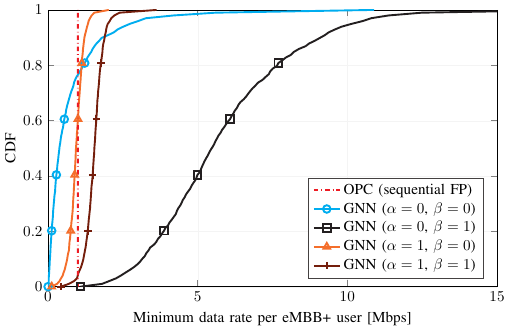}}
    \vspace{-2mm}
    \caption{CDF of the minimum rate of eMBB+ users vs. GNN training loss.}    
    \label{fig:10}
    \vspace{-4mm}
\end{figure}

Figs.~\ref{fig:9} and \ref{fig:10} present the EE and the minimum user rate of the GNN-based implementation w.r.t. different training losses, respectively. One can confirm that the ML algorithm developed in Section~\ref{sec:6} behaves well, since it accurately approximates the EE of the sequential FP technique (especially the lower tail, which is the goal of the maximin operation). However, as stated earlier, considering the conventional loss $\mathcal{L}({\alpha})$ that focuses solely on the power coefficients and/or EE (cf. \eqref{eq:42}), does not ensure constraints $C1-C5$ are satisfied. In particular, we observe that for $\beta = 0$ (no augmented Lagrangian), more than $60\%$ of the time users do not meet the QoS. On the contrary, when introducing the additional penalty (i.e., $\beta = 1$), this percentage is significantly reduced and becomes (practically) zero. Remarkably, in this specific setup, setting $\alpha = 0$ and $\beta = 0$ yields the best outcome: the GNN closely matches the analytical solution, and all users safely meet the rate constraint.

Finally, recall that the GNN is not only trained for a single CF deployment but can, given its generalization properties, handle diverse scenarios. These are illustrated in Table~\ref{tab:2}, where we present the Kullback-Leibler (KL) divergence, which measures the similarity between the two distributions of the EE (analytical solution and GNN prediction) \cite{Boy04}. For clarity, the baseline configuration ($K_u = 2$, $K_d = 10$, $M = 10$, $M_s = 5$) is marked in bold. As can be inferred, values close to zero further underscore the accuracy observed in the previous figures. On the other hand, for the maximin performance, we also include the the $95\%$-likely loss, defined as the relative error at the bottom $5$-th percentile (that is, the coverage quality for $95\%$ of the users). Overall, regardless of the setting, we obtain a tight approximation (with all losses around $1\%$). These results then justify the use of GNNs for substituting expensive model-based solutions. 

\begin{table} [t]
    \caption{GNN performance for different scenarios.}
    \centering
    \begin{tabular}{|c|c|c|c|c|c|c|}
        \hline
        $K_u$ & $K_d$ & $M$ & $M_s$ & KL divergence & $95\%$-likely loss ($\%$)  \\
        \hline        
        $\mathbf{2}$ & $\mathbf{10}$ & $\mathbf{10}$ & $\mathbf{5}$ & $\mathbf{0.099113}$ & $\mathbf{0.19714}$ \\
        $2$ & $10$ & $10$ & $1$ & $0.13999$ & $3.0727$ \\
        $2$ & $10$ & $5$ & $5$ & $0.089932$ & $0.37763$ \\        
        $1$ & $10$ & $10$ & $5$ & $0.10854$ & $0.29992$ \\
        $1$ & $5$ & $10$ & $5$ & $0.059146$ & $1.7683$ \\
        \hline
    \end{tabular}        
    \label{tab:2}    
\end{table}

\subsection{Complexity Analysis}
Among other reasons, convex problems are relevant because they can be solved in polynomial time \cite{Boy04}. More precisely, it can be shown that their cost scales polynomially with the problem size (usually characterized by the number of variables and constraints). Since interior-point methods are used to solve the inner problem \eqref{eq:34} (cf. \cite{CVX24}), the complexity per Dinkelbach iteration of the sequential FP power control (i.e., line 5 in Algorithm~\ref{alg:2}) is $\mathcal{O}((K_u + K_d)^{3.5}\log(\epsilon^{-1}))$, with $\epsilon$ the desired numerical accuracy. This reflects the cubic cost of solving the associated linear system at each step of the interior-point procedure. In that sense, considering also the number of outer iterations (which depend on the convergence thresholds $\theta_0$ and $F_0$ established in line 2), the total complexity would be $\mathcal{O}(N_{\text{S}}N_{\text{D}}(K_u + K_d)^{3.5}\log(\epsilon^{-1}))$, where $N_{\text{S}}$ and $N_{\text{D}}$ are the number of iterations of the sequential FP (line 3) and Dinkelbach (line 4) loops, respectively.

On the other hand, once the learning tasks are completed offline (i.e., generating the database by solving the previous optimization problem and training the GNN), the complexity of the data-driven method will depend on the specific layout. In particular, for our heterogeneous GNN with $M (K_u + K_d)$ fully-connected nodes (either via active or inactive edges), the computational complexity is (mainly) given by the message passing function \cite{Mis24}, whose dominant operation is the edge-wise feature transformation and aggregation, i.e., computing messages along edges and summing them (cf. \eqref{eq:40}). Consequently, the complexity is governed by the number of edges, that is $\mathcal{O}(M(K_u + K_d)(M + K_u + K_d))$. Recall that this corresponds to the inference phase, in which the prediction is obtained in a single forward pass. Hence, since the cost is considerably lower than that of the analytical counterpart, one can conclude that GNNs are a feasible, low-complexity solution for power control in CF-mMIMO systems.

\section{Conclusions} \label{sec:8}
This paper has investigated the UL coexistence of eMBB+ and mMTC+ in terminal-centric CF-mMIMO systems. A non-orthogonal MA scheme based on time–frequency spreading for mMTC+ transmissions has been proposed to enable efficient sharing of the resource grid with eMBB+ traffic. Closed-form analytical expressions for the achievable data rate have been derived under imperfect CSI, incorporating FBL analysis to capture the short-packet nature of mMTC+. A power-control framework has then been developed to maximize the minimum EE of mMTC+ devices while meeting the QoS requirements of eMBB+ users. The resulting problem is solved via sequential FP, and a GNN-based solution is proposed to approximate the model-based approach with significantly lower complexity. Numerical results demonstrate the effectiveness of the proposed coexistence strategy and show that the GNN achieves near-optimal performance with a feasible computational cost.

\bibliographystyle{IEEEtran}
\bibliography{references}

\end{document}